\documentclass[12pt]{article}
\pdfoutput=1
\usepackage{jcappub}
\usepackage{feynmp}	
\usepackage{subfig}

\title{On the Expected Production of Gravitational Waves During Preheating }
\author{C. Armendariz-Picon}
\affiliation{Physics Department\\St.~Lawrence University, \\23 Romoda Dr., Canton, NY 13617, USA.}                                            
\emailAdd{carmendarizpicon@stlawu.edu}
\makeatletter
\def\@fpheader{\relax} 
\makeatother

\abstract{The existing predictions of the energy density of gravitational waves  produced during preheating  mostly rely on computer simulations in which the matter field inhomogeneities essentially behave classically. In this article we follow instead  a full quantum treatment of the process within the $in$-$in$ formalism. We use this approach to determine the expected density of the produced gravitational waves and numerically estimate its value in a simple scalar field model, neglecting backreaction. Particular attention  is devoted to the regularization and renormalization of the divergences that appear  in the $in$-$in$ formalism. We also address how our approach  compares with the conventional calculations used in the literature, and what elements could be missed in the conventional analyses of gravitational wave production that rely on numerical simulations. In the cases in which parametric resonance is effective, our results agree with the  predictions expected from  numerical simulations, as anticipated. At sufficiently low values of the resonance parameter, however, numerical simulations fail,  while our approach remains applicable.}

\begin{document}
\begin{fmffile}{graphs}

 \vspace{-5cm}
 
\maketitle

\section{Introduction}

The first direct detections of gravitational wave radiation by the LIGO and VIRGO collaborations \cite{Abbott:2016blz,Abbott:2016nmj,Abbott:2017vtc,Abbott:2017gyy,Abbott:2017oio,TheLIGOScientific:2017qsa} have opened a new window into our universe. The  reach of electromagnetic radiation is essentially limited  to the time of recombination, but the extremely weak interactions of gravitational waves allow them to propagate   practically  undisturbed since the early dawn of the Big Bang. Gravitational waves may thus allow us to probe processes that would remain unreachable otherwise, but  for that purpose, it is  imperative for quantitatively accurate predictions of the produced gravitational waves to be available. 
 
A prominent example of such processes is ``preheating" \cite{Kofman:1997yn}.  In order for the universe to transition from inflation to the radiation-dominated epoch of the Big Bang, the inflaton  needs to decay into matter at the end of inflation. Under the appropriate circumstances, the  inflaton to matter couplings responsible for its decay, along   with the oscillations of the inflaton  around the minimum of its potential,  lead to a stage of parametric amplification of the matter fields \cite{Kofman:1997yn,Allahverdi:2010xz,Amin:2014eta}. This strong amplification has the potential to source a background of gravitational waves that according to previous estimates may be detectable today \cite{Easther:2006gt,Easther:2006vd}. The literature on gravitational wave backgrounds is very extensive, and we just refer the reader to the reviews \cite{Guzzetti:2016mkm,Caprini:2018mtu} for further details. 

Previous analyses of gravitational wave production during preheating go back to the seminal article by Khlebnikov and Tkachev \cite{Khlebnikov:1997di}.  In that reference, the authors interpreted parametric amplification of the matter fields during preheating as the appearance of classical inhomogeneities, which were then responsible for  sourcing   gravitational waves  according to Einstein's quadrupole formula. Since then,  predictions of the expected energy density of the gravitational waves have relied essentially on lattice codes that simulate the evolution of these classical inhomogeneities in an expanding universe, and use the equivalent of Einstein's formula to calculate the gravitational wave spectrum \cite{Felder:2000hq,Frolov:2008hy,Easther:2010qz,Huang:2011gf}. One advantage of such approaches is that the backreaction of the matter fields on the evolution of the inflaton is readily taken into account. 

Yet the matter fields responsible for the generation of the gravitational waves during preheating do not begin in a classical state, but are assumed to be instead in the $in$ vacuum.  The justification for a classical analysis rests on  the heuristic argument that parametric resonance can be interpreted  as the production of large number of particles, and that modes with large occupation numbers essentially behave like classical waves \cite{Khlebnikov:1997di,Kofman:1997yn}.  The author is unaware of any  rigorous proof of such claims, and even if they did apply,  we would  certainly expect the classical approximation to fail as one  begins to probe parameters for which parametric resonance is  ineffective.  Part of the motivation of our study is to assess the regime in which  the classical approximation is appropriate. In particular, to the extent that a classical treatment of gravitational production  is supposed to be just  an approximation, the question arises as to what  it is exactly that one is trying to approximate. 

As we shall discuss, the spectral  energy density of gravitational waves is proportional to the square of their amplitudes. Hence,  one could argue that previous analyses  are just trying to estimate the power spectrum of the sourced gravitational waves.  Our main thesis is that the ``exact" expected  power spectrum ought to be computed using the now standard $in$-$in$ formalism \cite{Weinberg:2005vy}.  In that sense, gravitational wave production during preheating is conceptionally identical to that of of gravitational wave production during inflation, or that of the generation of  primordial scalar perturbations. The only difference is that, whereas the latter  just require  the evaluation of a tree-level diagram, the former involves a  one-loop diagram in which the matter fields that undergo parametric amplification run inside the loop (see figure \ref{fig:FeynmanCon}.) The two cubic vertices that appear in such a diagram just capture that gravitational waves are sourced by the energy-momentum tensor,  which is quadratic in the matter field at lowest order. For simplicity we do not take into account the backreaction of the matter fields on the evolution of the inflaton, although it ought to be possible to include it within the $in$-$in$ formalism.  

Our analysis  not only presents then an alternative and arguably simpler way of computing the predicted spectral density of  gravitational waves when backreaction is unimportant, but   perhaps more importantly, offers a well-defined framework to deal with the ultraviolet  divergences that appear in both kinds of approaches. Without a proper treatment of such  divergencies,  it is  not possible to extract sensible predictions from the underlying theory,  or to relate its predictions to properly renormalized parameters in the action of the theory.

\section{Framework}

We are going to study gravity coupled to an inflaton field $\phi$ that decays into a  scalar $\chi$ at the end of inflation,
\begin{equation}
S= \int d^4 x \sqrt{-g} \left[\frac{M_P^2}{2} R -\frac{1}{2}\partial_\mu\phi \partial^\mu \phi
 	-\frac{1}{2}\partial_\mu\chi \partial^\mu \chi-\frac{1}{2}m_\phi^2\phi^2
 	-\frac{1}{2}M_\chi^2\chi^2-\frac{\lambda}{2}\phi^2\chi^2\right].
\end{equation} 
For simplicity we consider a renormalizable quartic coupling between the inflaton and the scalar $\chi$, although our treatment could be easily  extended to a cubic coupling $\phi\chi^2$, or even to derivative interactions. The inflaton potential does not need to remain quadratic throughout field space, but only in the vicinity of its minimum at $\phi=0$. We are framing our analysis in the context of preheating after inflation, although our results ought to apply as well to any scenario in which an appropriately coupled massive scalar oscillates at the minimum of its quadratic potential. 

Gravitational waves are  represented by transverse and traceless metric perturbations $h_{ij}(t,\vec{x})$ of the background Friedman-Robertson-Walker metric,
\begin{equation}
	ds^2=a^2(t)\left[-dt^2+(\delta_{ij}+h_{ij}) dx^i dx^j\right].
\end{equation}
Note that the time coordinate $t$ is conformal time, in spite of the unusual label.
It shall prove useful to decompose these metric perturbations into components that do not mix under  translations and rotations,
\begin{equation}\label{eq:h expansion}
	h_{ij}(t,\vec{x})=\frac{1}{\sqrt{V}}\sum_{\vec{p},\sigma}h_\sigma(t,\vec{p}) Q_{ij}{}^\sigma(\vec{p})e^{i\vec{p}\cdot\vec{x}},
\end{equation}
where the $Q_{ij}{}^\sigma$ are appropriate polarization tensors (see appendix \ref{sec:Polarization Tensors}). Under  spatial translations  by an amount $\vec{T}$, $h_\sigma(t,\vec{p})$ changes by a phase factor $e^{i\vec{p} \cdot \vec{T}}$, and under a rotation by an angle $\theta$ about the $\vec{p}$ axis it changes by $e^{-i\sigma \theta}$. We work in a universe of finite volume ${V=L^3}$ and impose periodic boundary conditions on all the fields. Hence, the sum in equation (\ref{eq:h expansion}) runs over modes with $\vec{p}=\frac{2\pi}{L} \vec{n}$,  where $\vec{n}\in \mathbb{Z}^3$, and over the two possible helicities of a gravitational wave, $\sigma=\pm 2$.  The reality of the metric (and hence $h_{ij}$) implies that ${h_\sigma(t,\vec{p})=h_\sigma^*(t,-\vec{p})}$. 

\subsection{Couplings to Matter}
In order to compute the spectrum of gravitational waves produced during preheating, we need to determine how gravitational waves couple to matter. As we shall see, at next to  leading order in an expansion in $M_P^{-1}$ it suffices to consider couplings linear and quadratic in the graviton. At this order the interaction part of the action is 
\begin{equation}\label{eq:SI}
	S_I\equiv-\int dt\, (\mathcal{H}^{1}_I+\mathcal{H}_I^2)\equiv \int dt\, \sum_{\vec{p} ,\sigma}   
	\Big[S_1^\sigma(t,\vec{p}) h_\sigma(t,-\vec{p})+ S^\sigma_2(t,\vec{0}) h_\sigma(t,\vec{p}) h_\sigma(t,-\vec{p}) \Big],
\end{equation}
where we have restricted the  couplings quadratic in $h_\sigma$ to those with opposite momenta, which are the only ones we shall need.  The last equation implies that   the source  term $S_1^\sigma(t,\vec{p})$ is
 \begin{equation}
	S_1^\sigma(t,\vec{p})\equiv \frac{\delta S_m}{\delta h_\sigma(t,-\vec{p})}\bigg|_{h=0}=\frac{a^6}{2\sqrt{V}}\int d^3 x\, T^{ij} Q_{ij}{}^\sigma(-\vec{p}) \,e^{-i\vec{p}\cdot\vec{x}},
\end{equation}
where the $T^{ij}$ are the spatial components of  the energy-momentum tensor of matter in the background spacetime. Hence, as expected, the source of linearized gravity is the energy-momentum tensor. Restricting our attention to that of the matter field $\chi$ we find 
\begin{equation}\label{eq:S1}
	S_1^\sigma(t,\vec{p})=-\frac{a^2}{2\sqrt{V}}\sum_{\vec{k}_1, \vec{k}_2}
		\vec{k}_1\cdot \hat{\epsilon}^{\frac{\sigma}{2}}(-\vec{p}) \, \,
		\vec{k}_2\cdot \hat{\epsilon}^{\frac{\sigma}{2}}(-\vec{p})\, 
		\chi_{\vec{k}_1}\!(t) \chi_{\vec{k}_2}\!(t) \,
		\delta_{\vec{k}_1+\vec{k}_2,\vec{p}},
\end{equation}
where we have also expanded the matter fields in Fourier modes,
\begin{equation}
	\chi(t,\vec{x})\equiv \frac{1}{\sqrt{V}}\sum_{\vec{k}} \chi(t,\vec{k}) e^{i\vec{k}\cdot \vec{x}}
	\equiv \frac{1}{\sqrt{V}}\sum_{\vec{k}} \chi_{\vec{k}}(t) e^{i\vec{k}\cdot \vec{x}},
\end{equation}
and the polarization vectors $\hat{\epsilon}$ are those in appendix \ref{sec:Polarization Tensors}.  Note that only the gradient terms in the energy-momentum tensor source gravitational waves and that the  Kronecker delta enforces momentum conservation. The coupling quadratic in  gravitational waves is 
\begin{equation}\label{eq:S2}
	S_2^\sigma(t,\vec{0})=\frac{a^2}{8V}\sum_{\vec{k}}\left\{\left(m_0^2 a^2+\vec{k}\,^2-
	4| \vec{k}\cdot \hat{\epsilon}^{\frac{\sigma}{2}}(\vec{p})|^2\right)\chi(t,\vec{k})\chi(t,-\vec{k})-\dot{\chi}(t,\vec{k})\dot{\chi}(t,-\vec{k})\right\},
\end{equation}
where an overdot denotes a derivative with respect to conformal time $t$, and we have introduced the effective squared mass
\begin{equation}\label{eq:m0}
	m^2_0\equiv M_\chi^2 +\lambda \bar{\phi}^2.
\end{equation}
We denote the  background value of the inflaton by $\bar{\phi}$.

\subsection{Energy Density of Gravitational Waves}
Following Isaacson \cite{Isaacson:1968zza} we define the energy density of gravitational waves to be
\begin{equation}
	\rho\equiv \frac{M_P^2}{4a^2}\left[\sum_{ij}\dot{h}_{ij}(t,\vec{x})\dot{h}_{ij}(t,\vec{x})			\right]_\mathrm{avg}
	\!\!\!\!                             \approx \frac{M_P^2}{4a^2}
	\frac{1}{T_\mathrm{avg} V}
	\int dt \,
	\sum_{\sigma,\vec{p}}
	\dot{h}_{\sigma}(t,\vec{p}) \dot{h}_{\sigma}(t,-\vec{p}),
\end{equation}
 where we have neglected  derivatives of the scale factor (a good approximation in the short-wavelength limit),  and $[\,]_\mathrm{avg}$ indicates an average over a sufficiently large spacetime region of comoving size $T_\mathrm{avg} V_\mathrm{avg}$. To express the energy density in Fourier space, we have used  equation (\ref{eq:h expansion}) and assumed that the spatial volume of the region over which we average approaches the size of the finite universe,  $V_\mathrm{avg}\to V$. In that case, the spatial average picks up Fourier components of opposite momenta.

Our theories do not allow us to directly predict the energy density of gravitational waves in our particular universe, but, instead, they make statements about the  average energy density across an appropriate ensemble of universes.  In the quantum theory, the latter simplifies if we use that at short wavelengths the time derivative of a gravitational wave is proportional to $\omega_p=p$. Hence, the expected energy density in gravitational waves is 
\begin{equation}\label{eq:rho}
\langle \rho\rangle\equiv \int \frac{dp}{p} \frac{d\langle\rho\rangle}{d\log p},
\quad
\frac{d\langle\rho\rangle}{d\log p}=
\frac{M_P^2}{8\pi^2a^2} \sum_{\sigma}
\frac{p^5}{T_\mathrm{avg}}
\int dt \, 
\langle h_{\sigma}(t,\vec{p}) h_{\sigma}(t,-\vec{p})\rangle,
\end{equation} 
where $\langle\,\rangle$ denotes quantum-mechanical expectation, we have introduced the spectral density $d\rho/d\log p$, and approximated  the mode sum by an integral. The time interval of the average $T_\mathrm{avg}$ in equation (\ref{eq:rho}) is set to be several times the frequency of the  wave in the domain of interest, $p\, T_\mathrm{avg}\gg 1$. Typically, predictions of the energy density of gravitational waves are cast in terms of the fraction of the critical density, the so-called density parameter $\Omega_{GW}$, which in terms of the spectral  density becomes
\begin{equation}
\Omega_{GW}(p) \equiv \frac{1}{3 M_P^2 H^2}\frac{d\langle\rho\rangle}{d\log p},
\end{equation}
where $H$ is Hubble's constant at the time of interest.

\section{Gravitational Wave Production in the In-In Formalism}

Equation (\ref{eq:rho})  implies that in order to determine the energy density of gravitational waves all we need to compute  is the  gravitational wave power spectrum
\begin{equation}\label{eq:Pdef}
	P_\sigma(t,\vec{p})\equiv \langle h_{\sigma}(t,\vec{p}) h_{\sigma}(t,-\vec{p})\rangle,
\end{equation}
where the operators $h_\sigma(t,\vec{p})$ are in the Heisenberg picture.  In the $in$-$in$ formalism \cite{Weinberg:2005vy}, the expectation value (\ref{eq:Pdef}) can be expanded in different powers of the interaction. At zeroth order  the power spectrum is of order $M_P^{-2}$, because the properly normalized mode functions of the graviton are proportional to $M_P^{-1}$ (see below.) The leading   correction  is then of order $M_P^{-4}$, and arises from terms with at most two interaction vertices,
\begin{align}\label{eq:vev}
	&P_\sigma(t,\vec{p}) = \langle h_{\sigma}(t,\vec{p}) h^*_{\sigma}(t,\vec{p})\rangle
	+\int^{t}\!\!  d\bar t_1 \! \! \int^t \!\!  dt_1   \langle \mathcal{H}^{(1)}_I(\bar t_1) h_{\sigma}(t,\vec{p}) h^*_{\sigma}(t,\vec{p}) \mathcal{H}^{(1)}_I(t_1)\rangle
	\nonumber \\ 
	& - \!\! \int^{t} \!\! \! dt_1 \!\! \int^{t_1} \!\!\!\!  dt_2\, \langle  h_{\sigma}(t,\vec{p}) h^*_{\sigma}(t,\vec{p})\mathcal{H}^{(1)}_I(t_1)\mathcal{H}^{(1)}_I(t_2) \rangle 
	\! - \!\! \int^{t} \!\!  d\bar t_1 \!\!\int^{\bar t_1} \!\! \!\!d\bar t_2\,
	\langle \mathcal{H}^{(1)}_I(\bar t_2)\mathcal{H}^{(1)}_I(\bar t_1) h_{\sigma}(t,\vec{p}) h^*_{\sigma}(t,\vec{p}) \rangle 
	\nonumber \\
	&-i\int^{t} dt_1 \langle h_{\sigma}(t,\vec{p}) h^*_{\sigma}(t,\vec{p})
	\mathcal{H}_I^{(2)}(t_1)\rangle
	+i\int^{t} d\bar t_1 \langle \mathcal{H}_I^{(2)}(\bar t_1) h_{\sigma}(t,\vec{p}) h^*_{\sigma}(t,\vec{p})
	\rangle.
\end{align}
In this equation, all operators   are in the interaction picture, that is, evolve like free fields.  The time contours need to be chosen to project the vacuum of the free theory into that of the interacting theory in the asymptotic past. This is why we label the integration variables for in the expansions of $\langle 0| U_I(t)^\dag$ and $U_I(t)|0\rangle$ differently \cite{Adshead:2009cb}.  

\begin{figure}
\begin{center}
\begin{fmfgraph}(240,100) 
\fmfleft{h1} 
\fmfright{h2}
\fmfdot{h1,h2}
\fmf{wiggly}{h1,v1}
\fmf{plain,left,tension=0.5}{v1,v2,v1}
\fmf{phantom}{v2,v3}
\fmf{plain,right,tension=0.5}{v3,v4,v3}
\fmf{wiggly}{v4,h2}
\end{fmfgraph}
\caption{\label{fig:FeynmanDisc} Disconnected corrections to the power spectrum of gravitational waves in the $in$-$in$ formalism. Wavy and solid lines respectively represent gravitons and matter fields. Recall that each vertex is of the two possible types ``L" or ``R". } 
\end{center}
\end{figure}
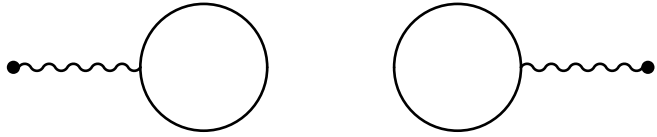

The first term on the right hand side of equation (\ref{eq:vev}) captures the vacuum fluctuations of the free gravitational field. It does not contain any information on the evolution of matter during preheating, so it does not reflect their production after inflation. We shall thus ignore this term during most of our analysis, although measurements  may be actually  able to distinguish  this tree-level contribution from the rest because of their different dependence on the momentum $\vec{p}$.    The remaining terms give rise to the Feynman diagrams shown in figures \ref{fig:FeynmanDisc} and \ref{fig:FeynmanCon}, and do depend on the evolution of the matter fields. These are the contributions that capture the production of gravitational waves by the amplified matter fields.  Although we have derived equation (\ref{eq:vev}) within the Hamiltonian formalism, the same expression would follow from the path integral. We will implicitly resort to the latter when renormalization forces us to introduce counterterms with derivatives of the metric fields.

\subsection{Disconnected Component}
We shall begin our analysis by studying the disconnected component of the power spectrum, which is represented  by the Feynman diagram in figure \ref{fig:FeynmanDisc}. From equation (\ref{eq:vev}), or by direct calculation of the expectation of $h_\sigma$  we find
\begin{equation}\label{eq:semiclassical}
	\langle h_{\sigma_1}(t,\vec{p}_1)\rangle\langle h_{\sigma_2}(t,\vec{p}_2)\rangle=
	\int^{t} \! dt_1 \!  \int^{t} \!  dt_2 \, D_{\sigma_1 p_1}^R(t;t_1) D_{\sigma_2 p_2}^R(t;t_2) 
	\langle S_1^{\sigma_1}(t_1,\vec{p_1}) \rangle \langle S_1^{\sigma_2}(t_2,\vec{p}_2) \rangle,
\end{equation}
where $D^R_{\sigma p}$ is the retarded Green's function of the gravitational waves,
\begin{equation}
D^R_{\sigma p}(t_1,t_2)\equiv i \, \theta(t_1-t_2)\left\langle \left[ h_\sigma(t_1,\vec{p}) ,h_\sigma(t_2,-\vec{p})\right]\right\rangle.
\end{equation}
The right hand side of equation (\ref{eq:semiclassical}) is precisely the expression we would use to calculate  energy density of gravitational waves in linearized semiclassical   gravity, in which the energy-momentum tensor is replaced by its expectation value. Since $S_1$  is quadratic in the matter fields, its expectation $\langle S_1^\sigma\rangle$ does not necessarily vanish, even in the  vacuum. But translational invariance does demand that $ \langle S_1^{\sigma}(t_1,\vec{p})\rangle\propto \delta_{\vec{p}}$, which implies that the right hand side of equation (\ref{eq:semiclassical})  is  proportional to two Kronecker deltas, $\delta_{\vec{p}_1} \delta_{\vec{p}_2} $.  In that case one cannot properly speak of gravitational waves, because  the corresponding metric perturbations are spatially constant. A spatially constant  expectation of the metric perturbations   effectively amounts instead to a shift in the scale factor.

A non-vanishing disconnected component in the spectrum of gravitational waves at non-zero momenta can only appear  if translational invariance is somehow broken, say, if the state of matter  describes a non-zero  number of quanta of definite momenta.   Although translational invariance  is broken in our realization of our universe (or so it seems),  it is certainly not broken in the ensemble, because the vacuum state of the perturbations  is invariant under translations, and so is the Hamiltonian (this follows from the invariance of the background under the same transformations.) Translational invariance can be broken only if a measurement causes the vacuum state of the fields to be projected onto a non-invariant state.  In the absence of such a measurement, it is worthwhile emphasizing that no matter how effective parametric resonance is, it cannot change the fact that both the vacuum state and the interaction Hamiltonian are  invariant under translations. For this reason, we do not expect the disconnected component to contribute to the spectrum of gravitational waves produced during preheating. 

\begin{center}
\begin{figure}
\subfloat
{
\begin{fmfgraph}(200,100) 
\fmfleft{h1} 
\fmfright{h2}
\fmfdot{h1,h2}
\fmf{wiggly}{h1,v}
\fmf{wiggly}{v,h2}
\fmf{plain}{v,v}
\end{fmfgraph}
}
\quad
\subfloat
{
\begin{fmfgraph}(200,100) 
\fmfleft{h1} 
\fmfright{h2}
\fmfdot{h1,h2}
\fmf{wiggly}{h1,v1}
\fmf{wiggly}{v2,h2}
\fmf{plain,left}{v1,v2,v1}
\end{fmfgraph}
}
\caption{\label{fig:FeynmanCon}Connected corrections to the power spectrum of gravitational waves in the $in$-$in$ formalism at order $M_P^{-4}$. Wavy and solid lines respectively represent gravitons and matter fields. Recall that each vertex is of two possible types, ``$L$" or ``$R$". } 
\end{figure}
\end{center}

\subsection{Connected Component}

Just as translational invariance implies that the correlation $\langle h_{\sigma_1}(t,\vec{p}_1) h_{\sigma_2}(t,\vec{p}_2)\rangle $ is proportional to $\delta_{\vec{p}_1,-\vec{p}_2}$, rotational invariance implies that the latter has to be proportional to $\delta_{\sigma_1,\sigma_2}$. Moreover, invariance under parity implies that 
\begin{equation}
	\left\langle h_{+2}(t,\vec{p}_1) h_{+2}(t,\vec{p}_2)\right\rangle=\left\langle h_{-2}(t,\vec{p}_1) h_{-2}(t,\vec{p}_2)\right\rangle, 
\end{equation}
Therefore,  the energy density of gravitational waves is the sum of the two equal energy densities in each helicity state. In the following, we  hence choose $\sigma=+2$ and  drop the explicit reference to the helicity in all our formulas, $h_{\vec{p}}(t)\equiv h_{+2}(t,\vec{p})$.  This choice is inconsequential, as the $\sigma=-2$ helicity behaves exactly the same way.

In order to calculate the connected component of the power spectrum (\ref{eq:Pdef}) it shall prove useful to split it into different contributions, according to the combination of vertex types they contain,
\begin{subequations}\label{eq:contributions}
\begin{equation}
P_{+2}(t,\vec{p})\equiv P_{LL}+P_{LR}+P_{RR}.
\end{equation}
The factor $P_{LL}$ captures the contributions from the two diagrams in figure \ref{fig:FeynmanCon} that only involve ``$L$" vertices, that is,  those terms in equation (\ref{eq:vev}) in which the interaction $\mathcal{H}_I$ appears to the left of the product of free fields $h (t,\vec{p}) h (t,-\vec{p})$. The factor $P_{RR}$ captures the ones that only involve ``$R$" vertices, that is, those terms in which the  interaction appears to the right of $h (t,\vec{p}) h (t,-\vec{p})$. And finally, $ P_{LR}$ contains  those in which one interaction vertex appears to the left, and one to the right of the product.  Substituting equations  (\ref{eq:S1}) and (\ref{eq:S2}) into (\ref{eq:vev}) we hence obtain

\begin{align}
\begin{split}\label{eq:PRR}
P&_{RR}= \frac{1}{4(2\pi)^3}\Bigg[- \int^{t} \! \! \! dt_1 \!  \int^{t_1} \! \! \!  dt_2 \!  \int \! d^3 k\,  a_1^2 a_2^2 \,k^4 \sin^4\theta 
D_{p}(t;t_1) D_{p}(t;t_2) 
G_k(t_1; t_2) G_q(t_1; t_2)
\\
& +i\int^{t} \!\!\!dt_1 a_1^2 D^2_{p}(t;t_1) \!\!
	\int \!\!\! d^3 k \Big\{\!\left(m_\chi^2 a_1^2+(1-\sin^2 \theta)k^2\right)G_k(t_1;t_1)
	-\langle \dot{\chi}(t_1,\vec{k})\dot{\chi}(t_1,-\vec{k})\rangle\Big\}\Bigg],
\end{split}
\\
P_{LR}&=\frac{1}{4(2\pi)^3} \int^{\bar{t}_f} \!\!\!  d\bar{t}_1\!  \int^{t} \! \! \!  dt_1 \! \int  \! \!  d^3 k\,  a^2(\bar{t}_1) a^2(t_1) k^4 \sin^4\theta 
D^*_{p}(t;\bar{t}_1)  D_{p}(t;t_1) 
G_k(\bar{t}_1;t_1) G_q(\bar{t}_1; t_1), \label{eq:PLR}
\\
P_{LL}&=P_{RR}^* ,
\end{align}
\end{subequations}
where $\vec{q}\equiv\vec{p}-\vec{k}$, $\theta$ is the angle between $\vec{k}$ and $\vec{p}$,  and  we have introduced the correlation functions
\begin{equation}\label{eq:correlators}
	D_{p}(t_1;t_2)\equiv \left\langle h(t_1,\vec{p}) h(t_2,-\vec{p}) \right\rangle,
	 \quad
	G_k(t_1;t_2) \equiv  \langle \chi(t_1,\vec{k}) \chi(t_2,-\vec{k})\rangle.
\end{equation}
Because of rotational invariance, the latter only depend on the magnitude of the vectors $\vec{p}$ and $\vec{k}$.

\subsection{Mode Functions}

In order to quantize the theory and determine the correlators (\ref{eq:correlators}) we expand the fields into creation and annihilation operators as usual,
\begin{equation}
	\chi_{\vec{k}}(t)=a_{\vec{k}} \, w_k(t)+ a^\dag_{-\vec{k}} \, w^*_k(t), \quad
	h_{\vec{p}}(t)=a_{\vec{p}} \, u_p(t)+ a^\dag_{-\vec{p}} \, u^*_p(t).
\end{equation} 
It is then convenient to introduce the rescaled mode functions
\begin{equation}
	\tilde{w}_k\equiv a  w_k, \quad
	\tilde{u}_p\equiv a u_p,
\end{equation}
which  obey the mode equations 
\begin{equation}\label{eq:mode equations}
	\ddot{\tilde{w}}_k+\left(k^2+m_0^2 a^2 -\frac{\ddot{a}}{a}\right)\tilde{w}_k=0, 
	\quad
	\ddot{\tilde{u}}_p+\left(p^2-\frac{\ddot{a}}{a}\right)\tilde{u}_p=0,
\end{equation}
subject to the normalization conditions $\tilde{w}_k \dot{\tilde{w}}_k^*-\dot{\tilde{w}}_k \tilde{w}^*_k =(M_P^2/4)(\tilde{u}_p \dot{\tilde{u}}_p^*-\dot{\tilde{u}}_p \tilde{u}^*_p)=i$. In the regime in which the squared frequencies inside the parenthesis of equation (\ref{eq:mode equations})  are slowly varying, the mode functions  $w_k$ and $u_p$ can  be expressed in WKB form, with time-dependent frequencies that at leading order in the adiabatic expansion are 
\begin{subequations}\label{eq:WKB expansions}
\begin{align}
 u_{\vec{p}}(t)&=\frac{2}{a \,M_P \sqrt{2H_p}}\exp\left(-i\int^t_{t_0} \!  H_p\, dt_1\right),
 	\quad
 	H_p=p+\cdots \label{eq:h WKB}
	\\
 w_{\vec{k}}(t)&=\frac{1}{a \sqrt{2W_k}}\exp\left(-i\int^t_{t_0} \!  W_k\,dt_1\right),
 	\quad
	W_k=\sqrt{k^2+m_0^2 a^2}+\cdots.\label{eq:adiabatic W}
\end{align}
\end{subequations}
We expect these approximate solutions to hold  when the modes of interest are well within the horizon. The appearance of the normalization factor $M_P^{-1}$ in the graviton mode functions implies that their propagator is of order $M_P^{-2}$, which is why all the contributions in equation (\ref{eq:contributions}) are of order $M_P^{-4}$. 

 Note that the effective mass of the matter field $\chi$ in equation (\ref{eq:m0})  depends on the background value of the inflaton. We assume that the background inflaton and the background metric obey the  equations of motion
\begin{equation}
\ddot{\bar{\phi}}+2\mathcal{H}\dot{\bar{\phi}}+m_\phi^2 a^2 \bar{\phi}=0,
 \quad
\mathcal{H}^2=\frac{1}{6M_P^2}\left(\dot{\bar{\phi}}^2-m_\phi^2 a^2\bar{\phi}^2\right),
\end{equation} 
where $\mathcal{H}=\dot{a}/{a}$.  Hence, our current approach does not take into account the backreaction of the matter fields on the evolution of the inflaton, nor the backreaction on the evolution of the metric.

\section{Regularization and Renormalization}
\label{sec:Regularization and Renormalization}

It is quite obvious that the diagrams whose contribution are given by $P_{LL}$, $P_{RR}$ and $P_{RR}$ in equations (\ref{eq:contributions}) are divergent and thus require regularization and renormalization. Say,  if we impose a sharp cutoff at spatial momenta $k=\Lambda$ , the leading contribution to $P_{RR}$  in the ultraviolet is  
\begin{equation}\label{eq:approx UV behavior}
	P_{RR}\sim\int^t dt_1 D^2_p(t;t_1) \int d^3k  \frac{k^4 \sin^4 \theta}{k^3},
\end{equation}
which grows with the fourth power of the cutoff.  In arriving at this expression we have made the short-wavelength approximation $W_k\approx k$, integrated by parts over $t_2$, and assumed that  the time variable has a small imaginary component, $t\to (1-i\epsilon)t$. This slight  clockwise tilt of the integration contour eliminates the contribution of the asymptotic past to the time integral (a detailed calculation follows below.)  It is clear from the structure of equation (\ref{eq:approx UV behavior}) that the origin of the ultraviolet divergence lies in the $k^4$  factor  in the integrand, which originates from the derivative interaction between gravitons and matter fields in equation (\ref{eq:S1}).

Yet a cutoff is  not  the appropriate way to regularize the integral.    Since our starting point is a generally covariant theory, it is important that the regularization preserve diffeomorphism invariance.  As emphasized in \cite{Weinberg:2010wq}, in the context of cosmological perturbation theory  dimensional regularization is not particularly convenient either, specially when the mode functions of the matter fields are not explicitly known. We shall follow instead a  a generally covariant implementation of Pauli-Villars \cite{Pauli:1949zm}.  A similar method was also   proposed  in a cosmological context in  \cite{Weinberg:2010wq}. The idea is to introduce a set of $N$ minimally coupled  scalar matter regulator fields $\chi_r$ ($r=1,\ldots, N$) of mass $m_r$ and Grassmann parity $\sigma_r$. Fields of even parity, $\sigma_r=1$,  are bosonic, and fields of odd parity, $\sigma_r=-1$, are fermionic.  Strictly speaking, the non-triviality of the action for the Grassmann-odd fields demands that the latter appear in pairs, $\chi_r$ and $\bar{\chi}_r$,  although for notational simplicity we shall not make this explicit. What matters is that loop contributions from the Grassmann-odd fields have the opposite sign as their bosonic counterparts. In that sense, the latter resemble the Faddeev-Popov ghosts of gauge theories though their purpose here is to cancel the divergences that appear in the ultraviolet. After this has been accomplished, we shall decouple the regulators by  taking  their masses to infinity. The removal of the regulators renormalizes the coefficients of the appropriate terms in the action. The reader may want to skip what remains of  this slightly technical section and jump directly to subsection \ref{sec:Section Summary}, which quickly summarizes the relevant results  of what follows.

\subsection{Renormalization of $P_{RR}$}
We begin our discussion with the  regularization of $P_{RR}$, equation (\ref{eq:PRR}). It turns out that only $P_{LR}$ contributes to the effective energy density of the gravitational waves, but the renormalization  of $P_{RR}$ will help us to set the stage for the renormalization of the former.  

Because, by construction, the regulator fields couple to gravity like the original matter field $\chi\equiv\chi_0$, they can also run in the matter loop of figure \ref{fig:FeynmanCon}. Their contribution to the two-point function of gravitational waves parallels that of $\chi_0$,  the only difference being that fermionic loops are proportional to an additional minus sign,
\begin{equation}\label{eq:PRRreg}
\begin{split}
	&P_{RR}= \sum_{i=0}^N \sigma_i\Big\{\frac{-1}{{4(2\pi)^3} } 
	 \int d^3 k\,   k^4 \sin^4\theta \!\!
	 \int^{t} \! \! \! dt_1 \! 
	 \int^{t_1} \! \!\!  dt_2  \, a^2_1 a^2_2 \, 
	D_{p}(t;t_1) D_{p}(t;t_2) 
	G^i_k(t_1; t_2) G^i_q(t_1; t_2)
\\
	& + \frac{i}{4(2\pi)^3} \!\!\int^{t}\!\!\! dt_1 a_1^2 D^2_{p}(t;t_1)\!\!
	\int \!\!d^3 k \Big[\left(m_i^2 a_1^2+k^2 \!-\! 2k^2\sin^2 \theta \right)G^i_k(t_1;t_1)
	  \!-\! \langle \dot{\chi}_i(t_1,\vec{k})\dot{\chi}_i(t_1,-\vec{k})\rangle\Big]
	\Big\}.
\end{split}
\end{equation}
The correlator of the $i$-th regulator field is denoted by $G^i_k(t_1;t_2)\equiv \langle \chi_i(t_1,\vec{p}) \chi_i(t_2,-\vec{p})\rangle$, and we have switched the order of integration. 

The values of the masses $m_i$ are dictated by the requirement that the sum in (\ref{eq:PRRreg}) be finite. We shall introduce  a  cutoff at  spatial momenta  $k=\Lambda$ first, and then  determine under what conditions the mode integral converges as  $\Lambda\to\infty$.  To estimate how a given integrand depends on the  momentum cutoff  we shall consider an adiabatic expansion in the number of time derivatives.  We begin by noting that the frequency $W_k$ in the mode functions (\ref{eq:adiabatic W}) has the adiabatic expansion
\begin{equation}
	W_k=\omega_k+\frac{3}{8}\frac{\dot\omega^2_k}{\omega_k^3}-\frac{1}{2\omega_k}\frac{\ddot{a}}{a}-\frac{\ddot{\omega}_k}{4\omega_k^2}+\cdots, \quad
	\omega_k=\sqrt{k^2+a^2 m_i^2},
\end{equation}
where we have omitted terms with four or more derivatives. In the limit of large $k$ the  time integral  over $t_2$  in equation (\ref{eq:PRRreg}) itself can be expanded  adiabatically by repeated integration by parts,
\begin{equation}\label{eq:integral expansion}
\begin{split}
	 \int_{-\infty}^{t_1} dt_2 \,& e^{-i  \int_{t_0}^{t_2} W_k(t_3) dt_3}f(t_2)=
	e^{-i \int_{t_0}^{t_1} W_k(t_3) dt_3}
	\sum^N_{n=0} 
	\left(\frac{1}{iW_k(t_1)}\frac{d}{dt_1}\right)^n \left( \frac{f(t_1)}{-i W_k(t_1)}\right) \\
	&-\int_{-\infty}^{t_1} dt_2 \, e^{- \int_{t_0}^{t_2} W_k(t_3) dt_3} \frac{d}{dt_2}\left(\frac{1}{i W_k(t_2)}\frac{d}{dt_2}\right)^N
	\left(\frac{f(t_2)}{-i W_k(t_2)}\right),
\end{split} 
\end{equation}
where we have used that the boundary terms in the asymptotic past vanish (because of the $i\epsilon$ prescription.) Since each time derivative is accompanied by a factor of $W_k^{-1}\sim k$, each one reduces the degree of divergence of the mode integral by one, so we just need to consider a finite number of  derivatives to find the divergent pieces of the integral.  The dependence of the mode integral on the cutoff can be  determined  now by expanding the frequencies $\omega_q$ in powers of $p$,
\begin{equation}
	\frac{1}{\omega^i_q}\approx \frac{1}{\omega_k^i}\left(1+\frac{k \,p\cos \theta}{(\omega_k^i)^2}-\frac{p^2(k^2-3k^2 \cos^2 \theta+a^2 m_i^2)}{2(\omega_k^i)^4}+\cdots\right).
\end{equation}
Again, each additional power of $p$ is accompanied by a factor $1/\omega_k$, which lowers the degree of divergence of the mode integral by one. Since $\int d\theta \sin^5\theta \cos^n\theta$ vanishes for odd $n$, only quadratic terms in $p$ effectively appear in the expansion. The ensuing integrals then contain linear combinations of integrands of the generic form $k^n/(\omega_k)^{2m}$, which  in the limit $\Lambda\to\infty$  approach
\begin{equation}\label{eq:div integrals}
	\int_0^{\Lambda} \! dk\, \frac{k^n}{(k^2+a^2 m_i^2)^m}\to
	\begin{cases}
	{\displaystyle \frac{\Lambda^{1+n-2m}}{1+n-2m} - \frac{m \, a^2 m_i^2 \, \Lambda^{n-2m-1}}{n-2m-1}}+\cdots, 
	\quad &1+n-2m\neq 0, \\
	{\displaystyle \log \frac{\Lambda}{a\, m_i}}+\mathcal{O}(\Lambda^0), \quad &1+n-2m=0.
	\end{cases}
\end{equation}
 At zeroth order in time derivatives we thus find
\begin{align}\label{eq:PRRzero}
&P_{RR}^{(0)}=-\frac{i\,u^2_p(t)}{40(2\pi)^2}\sum_i \sigma_i \int^{t} dt_1 \, u_p^*{}^2 \Bigg[3\Lambda^4
	+\frac{11 p^2- 98a^2 m_i^2}{42}\Lambda^2 \\
	&-\frac{p^4+10\, p^2 m_i^2a^2-30\, a^4m_i^4}{12} \log\frac{2\Lambda}{a\, m_i}+\frac{192p^4+1570 p^2 a^2 m_i^2-2415 a^4 m_i^4}{2520}+\mathcal{O}\left(\frac{1}{\Lambda}\right)\Bigg] \nonumber, 
\end{align}
where all time-dependent functions  in the integrand (including $m_i^2$) are evaluated at time $t_1$.   Cancellation of the quartic, quadratic and logarithmic cutoff-dependent terms  in equation (\ref{eq:PRRzero}) hence requires that
\begin{equation}\label{eq:mi cond}
	\sum_{i} \sigma_i=0, \quad \sum_{i} \sigma_i M_i^2=0, 
	\quad \sum_{i} \sigma_i M_i^4=0,
\end{equation}
where we have assumed that the regulator fields couple to the inflaton just like the original field,  $m_i^2=M_i^2+\lambda \bar{\phi}^2$.  We shall  decouple the regulator fields by sending their masses $M_r$  to infinity.  The cancellation of the ultraviolet divergences  survives in this limit, but the presence of the logarithmic factors  implies that the dependence  on the masses $M_r$ persists and does not remain finite as the regulators are removed. These new divergences need to be renormalized by introducing  appropriate counterterms, as we shall discuss below.  This is a reflection of the conventional lore of low-energy effective field theory, namely, that the physics at high scales (the regulators) only affects low-energy observables through  the renormalization of the appropriate operators in the low-energy theory \cite{Appelquist:1974tg}.

\begin{figure}
\begin{center}
\begin{fmfgraph}(200,50) 
\fmfleft{h1} 
\fmfright{h2}
\fmfdot{h1,h2}
\fmf{wiggly}{h1,v}
\fmf{wiggly}{v,h2}
\fmfv{decoration.shape=cross}{v} 
\end{fmfgraph}
\end{center}
\caption{\label{fig:counterterm}Insertion of a quadratic counterterm needed to renormalize the power spectrum of gravitational waves. Here the vertex can also be of two types: $``L"$ and $``R"$.} 
\end{figure}
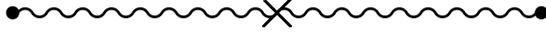

 We proceed next to higher orders in the expansion in time derivatives. At one time derivative,  the leading divergences are cubic and linear in the cutoff, with no logarithmic divergence,
 \begin{equation}
 \begin{split}
 &P_{RR}^{(1)}=\frac{u_p^2(t)}{12(2\pi)^2}\sum_i \!\sigma_i \! \int ^t \!\!\!dt_1 \!
 \Bigg[ \!-\!  \frac{u^*_p \dot{u}_p^*}{15}\Lambda^3
 \!+\frac{56a^2 (\mathcal{H}m_i^2+ m_i\dot{m}_i)u_p^*{}^2 \!+\!(13p^2+56a^2 m_i^2)u_p^*\dot{u}_p^*}{140} \Lambda \\
 &-\!\frac{96a^3(\mathcal{H} m_i^3+ m_i^2 \dot{m}_i)u_p^*{}^2
 \!+\! 9p^2 a(\mathcal{H} m_i+\dot{m}_i)u_p^*{}^2
 \!+\!(18 p^2 a m_i \!+\!64a^3 m_i^3)u_p^* \dot{u}_p^*}{256\pi^{-1}}\Bigg]_{t_1}
 \!\!\!\!\!+\mathcal{O}\left(\frac{1}{\Lambda}\right).
 \end{split}
 \end{equation}
Note the presence of a time derivative of $m_i$ in the linearly divergent term; it  can be eliminated upon integration by parts, which renders the integral convergent as $\Lambda\to \infty$, provided that  conditions (\ref{eq:mi cond}) are  satisfied. But inspection of the $\mathcal{O}(\Lambda^0)$ term in the integrand also reveals pieces that would diverge as the regulator masses $M_r$ are sent to infinity.  Because none of these terms is proportional to a power of the cutoff $\Lambda$, the only condition on their sum is that it remain finite in the limit $M_r\to \infty$. To determine the finite remainder we  would presumably need to impose   additional renormalization conditions, but  since there is no   counterterm with a single time derivative acting on the metric, its value  remains ambiguous in this approach. On the other hand,  if we had used dimensional regularization   to render integrals  of the form (\ref{eq:div integrals}) finite,  we would have set
 \begin{equation}
	 \int_0^\infty\! \! dk\,  \frac{k^n}{(k^2+a^2 m_0^2)^m}
	 \to(a \, m_0)^{1+n-2m}
	 \frac{\Gamma\left(\frac{1+n}{2}\right)}{2\Gamma(m)}
	 \Gamma\left(-\frac{1+n-2m}{2}\right).
 \end{equation}
Because the gamma function has poles only at negative integer values of its argument, odd positive powers of $a \,m_0$ would be proportional to the same   finite coefficient in Pauli-Villars regularization  if  we impose the additional conditions
\begin{equation}\label{eq:mi add cond}
	\sum_r \sigma_r M_r=0, \quad \sum_r \sigma_r M_r^3=0.
\end{equation}
Even powers of $a\,m_0$ would be multiplied with divergent coefficients as the limit of spatial dimensions approaches three, and  would require renormalization as before.

 The terms in the adiabatic expansion of equation  (\ref{eq:PRRreg}) that contain two time derivatives are  
\begin{align}\label{eq:PRRtwo}
P_{RR}^{(2)}=&-\frac{i}{240} \frac{u^2_p(t)}{(2\pi)^2}\sum_i \sigma_i \int^{t}  dt_1\Bigg[
\left\{14\mathcal{H}^2 u_p^*{}^2-16\dot{\mathcal{H}}u_p^*{}^2
+u_p^* \ddot{u}_p^* \right\}\Lambda^2
 \\
&+\Big\{5\left(p^2 (\mathcal{H}^2+\dot{\mathcal{H}})-2a^2m_i^2 ( \mathcal{H}^2+2\dot{\mathcal{H}})-4a^2 (\mathcal{H} m_i \dot{m}_i+\dot{m}_i^2+m_i\ddot{m}_i) \right) u_p^*{}^2
 \nonumber \\
&-10a^2\left(\mathcal{H} m_i^2+m_i\dot{m}_i\right)\dot{u}_p^* u_p^*
-\left(p^2+5a^2 m_i^2\right)u_p^* \ddot{u}_p^*\Big\}\log \frac{2\Lambda}{a m_i}
\Bigg]_{t_1}+\mathcal{O}\left(\frac{1}{\Lambda}\right), \nonumber
\end{align}
where, for simplicity, we have omitted the  terms of order $\Lambda^0$, which vanish because of conditions (\ref{eq:mi cond}), (\ref{eq:mi add cond}), or because the regulator masses approach infinity.   Again, the first equation in (\ref{eq:mi cond}) guarantees a finite limit as ${\Lambda\to \infty}$, but a logarithmic  dependence on the regulator masses $M_r$ survives the cancellation as the regulators are removed. Along the same lines we would find that there are no logarithmic divergences within the terms in $P_{RR}$ containing three time derivatives, and that those with four derivatives diverge logarithmically with the cutoff (and the regulator masses.) 

The logarithmic dependence on $M_r$ that signals the impact  at low momenta of the physics at  much higher scales can be reabsorbed into appropriate low-energy renormalized parameters. Inspection of the logarithmic divergences in equations (\ref{eq:PRRzero}) and (\ref{eq:PRRtwo})  reveals that the counterterms need to be operators with up to four derivatives of the metric.  Since our regularization respects diffeomorphism invariance, the former have to be of the form
\begin{equation}\label{eq:counterterms}
	S_c=\int d^4 x \sqrt{-g}\left[ c_0+c_2 R +c_{4a} R^2+c_{4b} R_{\mu\nu} R^{\mu\nu}\right],
\end{equation}
where we have used that in four spacetime dimensions the most general dimension four curvature invariant is a linear combination of $R^2$ and $R_{\mu\nu}R^{\mu\nu}$.
Noting that the expansion of the counterterm proportional to $c_0$ to quadratic order in the tensor modes  is
\begin{equation}
 \int d^4 x \sqrt{-g}= -\frac{1}{4} \int dt \, a^4 \sum_{\vec{p}} h_{\vec{p}}h_{-\vec{p}}+\mathcal{O}(h^3),
\end{equation}
and comparing the latter with equation (\ref{eq:PRRzero})  we realize that the term proportional to $a^4\sum_i \sigma_i m_i^4 \log m_i$  can be canceled by a single insertion of an  ``$R$" vertex  proportional to $c_0$, as in figure \ref{fig:counterterm}. In order to cancel the dependence  on the regulator masses, the counterterm coefficient has to be  
\begin{equation}\label{eq:c0}
	c_0=\frac{1}{8(2\pi)^2}\sum_r \sigma_r \left(M_r^2+\lambda \phi^2\right)^2 \log M_r+\mathrm{finite},
\end{equation}
which amounts to a renormalization of the cosmological constant and the inflaton potential. It is certainly not a coincidence that the $\phi$-dependent piece of this  counterterm is  precisely the one  needed to cancel the  divergences in the effective potential of the inflaton field due to its couplings to $\chi$.   As usual, the value of $c_0$ is   determined only up to  a finite  field-dependent constant, which needs to be fixed by appropriate renormalization conditions.   In appendix \ref{sec:Effective Equations of Motion} we discuss an example.

In conjunction with the term proportional to $10\, p^2 m_i^2 a^2$ in equation (\ref{eq:PRRzero}), comparison of equation (\ref{eq:PRRtwo}) with the expansion of the Einstein-Hilbert action to second order in the tensor modes of, say, positive helicity,
\begin{equation}
\int d^4 x\sqrt{g}R\supset\int dt \frac{a^2}{4}\sum_{\vec{p}} \left[-3\dot{h}_{\vec{p}}\dot{h}_{-\vec{p}}-4h_{\vec{p}}\ddot{h}_{-\vec{p}}-12\mathcal{H} h_{\vec{p}}\dot{h}_{-\vec{p}}
-(6\mathcal{H}^2 +6\dot{\mathcal{H}} +p^2) h_{\vec{p}} h_{-\vec{p}}
\right],
\end{equation}
allows us to find the regulator-dependent piece of the counterterm proportional to  the Ricci scalar $R$,\begin{equation}\label{eq:c2}
	c_2=-\frac{1}{24(2\pi)^2}\sum_r \sigma_r \left(M_r^2+\lambda\phi^2\right) \log M_r+\text{finite}.
\end{equation}
Thus, radiative corrections force us to introduce not only an Einstein-Hilbert term into the  action, but also a a non-minimal coupling of the inflaton to gravity.  Note that the radiative corrections we have considered arise only from the couplings of matter to gravity, and do not depend on the dynamics of gravity itself; the only assumption that does enter our analysis is that gravitational waves can be expanded in a set of creation/annihilation operators.

To arrive at the identification of the counterterm  (\ref{eq:c2}) we have discarded a total time derivative in $P^{(2)}_{RR}$ of the form
\begin{equation}
P^{(B)}_{RR}=\frac{iu_p(t)^2}{48(2\pi)^2}\sum_i \sigma_i \log \frac{\Lambda}{am_i} 
\int^t_{-\infty} dt_1 \frac{d}{dt_1}\left\{-2a^2 \mathcal{H} m_i^2 u_p^*{}^2-3a^2 m_i^2 u_p^*\dot{u}_p^*+2a^2 \dot{m}_i^2 u_p^*{}^2\right\}.
\end{equation}
In other words,  the log divergent pieces of $P_{RR}$ can be expanded as the sum of bulk and boundary terms, with the bulk contribution to  $P_{RR}$ being canceled by the counterterm proportional to $c_2$. The remaining piece is the boundary term above.  Because there is an analogous contribution from  the $LL$ diagram, which is just the complex conjugate of $P_{RR}$,  most of these boundary terms cancel.  There are however two terms that survive, namely, $P^{(B)}_{LL}+P^{(B)}_{RR}\propto 3ia^2m_i^2|u_p|^2(u_p\dot{u}_p^*-u^*_p\dot{u}_p)$, which happens to involve terms with time derivatives of the graviton mode functions.\footnote{This contribution is proportional to the Wroskian of the graviton mode functions, but since we prefer not to make any assumptions about the dynamics of gravity at this point, we shall leave it unevaluated. } Although it may appear strange at first that boundary terms contribute to the expectation of an observable $\mathcal{O}$, this has been previously noted  in the literature \cite{Arroja:2011yj}. In fact,  it is relatively easy to see that  boundary terms matter if they do not commute with  $\mathcal{O}$, as we discuss in appendix \ref{sec:Boundary Terms in the Interaction}.  Below we shall argue that the corresponding logarithmic divergence can be canceled by a counterterm proportional to the York-Hawking-Gibbons action.

We could proceed to determine the values of $c_{4a}$ and $c_{4b}$, by looking at the terms with four time derivatives in   equation (\ref{eq:PRRreg}). But at this point the algebra becomes increasingly involved, and we shall not need these counterterms anyway. In fact, the structure of the counterterm Lagrangian (\ref{eq:counterterms}) has been actually  known for a long time, at least since 't Hooft and Veltman's work on the one-loop divergences in gravity \cite{tHooft:1974toh}.  Within Pauli-Villars regularization, in the absence of spacetime boundaries, the required counterterms were discussed in reference \cite{Asorey:2003uf}. In the presence of spacetime boundaries, all the analyses known to the author  involve the heat kernel, which also demands the introduction of boundary terms to fully renormalize the effective action; see reference \cite{Vassilevich:2003xt} for a practical review.   The point of our analysis has been to illustrate that we can carry out the regularization and renormalization program  in a cosmological background, while preserving diffeomorphism invariance, using Pauli-Villars regularization.   The bulk divergences in the diagram with two $RR$ vertices are those that would be encountered in the standard $in$-$out$ calculations, such as those in \cite{tHooft:1974toh}. These divergences are only sensitive to the short-distance structure of the theory, and thus do not depend on the actual limits of integration. Although we have restricted our analysis to the diagrams with $R$ vertices, the same conclusion (and counterterms) would follow from  $P_{LL}$, which is simply the complex conjugate of $P_{RR}$.

\subsection{Renormalization of $P_{LR}$}

The only remaining contribution to the power spectrum is that  of $P_{LR}$, which  is the only one we shall actually need.  From  equation (\ref{eq:PLR}), by  including the regulator fields running in the loop, and inserting an ultraviolet cutoff for latter convenience, the latter reads
\begin{equation}\label{eq:PLRreg}
	P_{LR}= \frac{|u_p(t)|^2}{4(2\pi)^3}\sum_{i=0}^N \sigma_i\int\limits_{k\leq\Lambda} d^3 k \left| k^2 \sin^2 \theta  
	\int^{t} \! d\bar{t}_1a^2(\bar{t}_1) u_p(\bar{t}_1) w^i_k(\bar{t}_1) w^i_q(\bar{t}_1)\right|^2.
\end{equation}
We shall analyze $P_{LR}$ using a double expansion  in the number of time derivatives and powers of the external momentum $p$.  Up to three time derivatives the results are 
\begin{subequations}\label{eq:PLRexp}
\begin{align}
P_{LR}^{(0)}&\approx \sum_i \frac{|u_p(t)|^4 }{2(2\pi)^2} \sum\sigma_i 
\left[\frac{\Lambda^3}{90}- \left(\frac{13p^2}{840}+\frac{m_i^2 a^2}{15}\right)\Lambda +\frac{3\pi\, p^2  a\, m_i }{256}+\frac{\pi \, a^3\, m_i^3 }{24}\right], 
\\
P_{LR}^{(1)}&\approx    \frac{i|u_p|^2 \left(u_p \dot{u}^*_p-\dot{u}_p u_p^* \right)}{48(2\pi)^2}
 \sum_i \sigma_i
\left[\frac{\Lambda^2}{5}+\frac{277 p^2}{1050}+\frac{31 a^2 m_i^2}{30} -\left(\frac{p^2}{5}+a^2 m_i^2\right)\log \frac{2\Lambda}{am_i}\right],
\\
P_{LR}^{(2)}&\approx \frac{|u_p(t)|^2}{16(2\pi)^2} \sum_i  \sigma_i 
 \Bigg[
 \pi(u_p \dot{u}_p^*+\dot{u}_p u_p^*) \frac{\mathcal{H} a\, m_i}{32}
-\pi\dot{u}_p \dot{u}_p^* \frac{a\, m_i}{16}
+\pi(u_p \ddot{u}_p^*+\ddot{u}_p u_p^*)\frac{a\,m_i}{16}  \nonumber \\
&+\left(8|u_p|^2 \frac{\ddot{a}}{a}+|\dot{u}_p|^2-\ddot{u}_p u_p^*- u_p \ddot{u}^*_p\right)\frac{\Lambda}{15} 
+\pi |u_p|^2 a \,m_i  \left(\frac{19\mathcal{H}^2}{384}-\frac{5 \ddot{a}}{12a}\right)
\Bigg],
\\
P_{LR}^{(3)}&\approx   \frac{i|u_p|^2}{12(2\pi)^2} \sum_i  \sigma_i
\Bigg[
\left(\frac{u_p \dot{u}^*_p-\dot{u}_p u^*_p }{4}\frac{\ddot{a}}{a}+\frac{\dot{u}_p \ddot{u}^*_p-\ddot{u}_p \dot{u}^*_p-u_p \dddot{u}^*_p+\dddot{u}_p u^*_p}{40} \right)\log \frac{2\Lambda}{a m_i}
 \nonumber \\
&-\frac{u_p \dot{u}^*_p-\dot{u}_p u^*_p }{3}
\frac{\ddot{a}}{a}+\frac{u_p \ddot{u}^*_p-\ddot{u}_p u^*_p}{40}\mathcal{H}
-\frac{23\left( \dot{u}_p \ddot{u}^*_p-\ddot{u}_p \dot{u}^*_p - u_p \dddot{u}^*_p+\dddot{u}_p u^*_p \right)}{600}\Bigg], 
\label{eq:PLR3}
\end{align}
\end{subequations}
where we have omitted the pieces of order $\Lambda^{-1}$, and we list only those terms that do not vanish as $m_i\to \infty$. As in the case of $P_{RR}$ and $P_{LL}$, $P_{LR}$ remains finite in the limit $\Lambda\to \infty$ if conditions (\ref{eq:mi cond}) are satisfied. Since each subsequent derivative lowers the degree of divergence of  $P_{LR}$  by one power of the cutoff, we do not need to go beyond three derivatives.

But we still need to discuss how to remove the  dependence on the regulator masses that remains in equations (\ref{eq:PLRexp}) even when the cutoff is removed. As we  attempt to decouple the regulators,  new divergences  arise as the regulator masses approach infinity. These are of two types: i) Polynomial divergences, proportional to an  even number of time derivatives and no factor of  $i$, and ii) logarithmic divergences, proportional to  an odd number of time derivatives and a factor of $i$. Divergences of the first type  cancel because of conditions   (\ref{eq:mi add cond}). To eliminate the divergences of the second type 
it does  not suffice to add  generally covariant counterterms to the action,  as in equation (\ref{eq:counterterms}),  because the latter introduce ``bulk" corrections that involve  integrals over the whole spacetime, from $-\infty$ to $t$, rather than corrections at a single time $t$. Therefore, in order to remove the divergences of the second type  we shall include  an appropriate  boundary action in the theory. Diffeomorphism invariance  at the boundary suggests  that the required counterterms  should be constructed as a spatial integral over the boundary at $t$ of local invariants on that hypersurface.  At one time derivative it suffices to consider a counterterm proportional to the  York-Gibbons-Hawking boundary action \cite{York:1972sj,Gibbons:1976ue}
\begin{equation}\label{eq:b counterterm}
		c_1 \int d^3x \sqrt{\gamma}\,  K,
\end{equation}
where $\gamma_{ij}=g_{ij}$ is the three metric on the hypersurface at constant $t$, $K$ is the trace of the extrinsic curvature $K_{\mu\nu}=\nabla_\mu n_\nu$ and $n_\mu=-a\,\delta_{\mu 0}$ is the outward normal to the hypersurface.  Expanding the previous equation to second order in gravitons of positive helicity  we find
\begin{equation}
		\int d^3x \sqrt{\gamma}  \, K \supset -\frac{a^2}{2} \left(\dot{h}_{\vec{p}} h_{-\vec{p}}+\frac{3}{2}\mathcal{H} \, h_{\vec{p}} h_{-\vec{p}}\right).
\end{equation}
Therefore in the $in$-$in$ formalism, a counterterm of the form (\ref{eq:b counterterm}) would contribute two mutually conjugate corrections to the power spectrum: One in which the vertex in figure \ref{fig:counterterm} is an $L$ vertex, and one in which it is an $R$ vertex. Comparing the latter with the sum of the logarithmically divergent contributions in $P_{LL}$, $P_{RR}$ and  $P_{LR}$ with one derivative at the boundary,
\begin{equation}
-\frac{i |u_p|^2}{12 (2\pi)^2}\sum_i \sigma_i  \,a^2 m_i^2 (u_p \dot{u}_p^*-u^*_p \dot{u}_p)\log\frac{\Lambda}{a m_i},
\end{equation}
we can immediately read off the divergent piece of the corresponding coefficient,
\begin{equation}
	c_{1}=-2c_2.
\end{equation}
Remarkably, this relation between the two coefficients $c_1$ and $c_2$ is the same as the one originally proposed by York, Hawking and Gibbons in their boundary action \cite{York:1972sj,Gibbons:1976ue}. It is also the relation that emerges from heat kernel calculations of the effective action for both Dirichlet and Neumann boundary conditions on the fields \cite{Vassilevich:2003xt}. Note by the way that our analysis does not yield all necessary counterterms: If we had calculated the expectation of $\dot{h}_{\vec{p}} \dot{h}_{\vec{p}}$ we would have had to consider additional boundary counterterms constructed out of the effective (inflaton-dependent) masses of the matter fields.  Returning to the case at hand, we could proceed again  to determine the boundary counterterms needed to eliminate the terms with three time derivatives proportional to $\sum_i \sigma_i \log m_i$ in equation (\ref{eq:PLR3}) along the same lines, although we shall not do it here.

We are finally ready to compute the finite, renormalized value of $P_{LR}$. To do so we begin with the  identity
\begin{equation}\label{eq:PLR identity}
	P^\mathrm{ren}_{LR}\equiv\sum_{i=0}^n P_{LR}|_{i}+P^\mathrm{ct}_{LR}=P_{LR}\big|_{i=0}-P^{(\infty)}_{LR}\big|_{i=0}+P^{(\infty)}_{LR}|_{i=0}+\sum_{r=1}^n P_{LR}|_{r}+P^\mathrm{ct}_{LR},
\end{equation}
where $P_{LR}|_i$ denotes the contribution of the $i$-th field to $P_{LR}$,  $P_{LR}^{(\infty)}\big|_{i}$  the piece of that contribution that diverges with the cutoff, and $P_{LR}^\mathrm{ct}$ the contribution of the counterterms. Note that as both $\Lambda$ and the regulator masses tend to infinity  $P_{LR}|_r$ approaches the values that we collect in  (\ref{eq:PLRexp}). Now, as we remove the cutoff , $P_{LR}|_{i=0}-P^{(\infty)}_{LR}|_{i=0}$ remains finite by construction, and  the cutoff dependence of $P^{(\infty)}_{LR}|_{i=0}+\sum_{r=1}^n P_{LR}|_{r}$ cancels out because of equations (\ref{eq:mi cond}). This renders $ P^\mathrm{ren}_{LR}$ finite and cutoff-independent, but such a  regularized expression still depends on the mass of the regulators. As we decouple the latter,  even powers of  all the field masses in $P^{(\infty)}_{LR}|_{i=0}+\sum_{r=1}^n P_{LR}|_{r}$ cancel again because of (\ref{eq:mi cond}), but only odd powers of the regulator masses disappear, because of equation (\ref{eq:mi add cond}).  Factors that depend on the logarithm  of the regulator masses are rendered finite by the counterterms $P_{LR}^\mathrm{ct}$. Therefore,  the final, finite, renormalized value of $P_{LR}$ becomes
\begin{align}\label{eq:PLR final}
	&P^\mathrm{ren}_{LR}=\lim _{\Lambda\to\infty}\frac{|u_p(t)|^2}{(2\pi)^2}\Bigg\{\frac{1}{4}\int_0^\pi d\theta \sin^5 \theta \int_0^\Lambda dk\,   k^6  \left|
	\int\limits_{-\infty}^{t_f} \! d\bar{t}_1a^2(\bar{t}_1) u_p(\bar{t}_1) w_k(\bar{t}_1) w_q(\bar{t}_1)\right|^2\\
	&-\frac{1}{30}\left[|u_p|^2
\left(\frac{1}{6}\Lambda^3- \frac{13p^2}{56} \Lambda- a^2 m_0^2 \, \Lambda\right)\right]_{t_f}
 \nonumber \\
&- \frac{i}{240}\left(u_p \dot{u}^*_p- \dot{u}_pu_p^*\right)\left[\Lambda^2+
\frac{277 p^2}{210}+\frac{31 a^2m_0^2}{6}
-p^2\log\frac{\Lambda}{a \mu_1}  -5 a^2 m_0^2\log\frac{\Lambda}{a \mu_2} \right]_{t_f}
\nonumber \\
&- \frac{1}{240} 
\left[ \left(8|u_p|^2 \frac{\ddot{a}}{a}+|\dot{u}_p|^2-\ddot{u}_p u_p^*-u_p \ddot{u}^*_p \right)\Lambda\right]_{t_f}
\nonumber \\
& +\frac{i}{12}\left[\frac{u_p \dot{u}^*_p- \dot{u}_pu_p^*}{3}\frac{\ddot{a}}{a}-
\frac{u_p \ddot{u}^*_p- \ddot{u}_pu_p^*}{40}\mathcal{H}
+\frac{23( \dot{u}_p \ddot{u}^*_p-\ddot{u}_p \dot{u}^*_p-u_p \dddot{u}^*_p+\dddot{u}_p u^*_p)}{600}\right]_{t_f}
\nonumber \\
&- \frac{i}{480}\left[10 \frac{\ddot{a}}{a} (u_p \dot{u}^*_p \!-\! \dot{u}_p u^*_p)\log \frac{\Lambda}{a \mu_3}+ (\dot{u}_p \ddot{u}^*_p \!- \! \ddot{u}_p \dot{u}^*_p)\log \frac{\Lambda}{a \mu_4}-(u_p \dddot{u}^*_p\!-\!\dddot{u}_p u^*_p)\log \frac{\Lambda}{a \mu_5}\right]_{t_f}\Bigg\}. \nonumber
\end{align}

This is one of the main results of this article.  $P^\mathrm{ren}_{LR}$ converges because the cutoff dependence of the  integral is subtracted out, thus rendering the limit finite at the same time.  What does remain is the dependence on the arbitrary parameters $\mu_i$, which capture the ambiguities in the finite part of the  counterterms. Since we have not explicitly determined all of the latter, it is possible for some of the $\mu_i$ to be related to each other. Their specific values  can be fixed by imposing appropriate renormalization conditions. Say, in the renormalization scheme we discuss in appendix \ref{sec:Effective Equations of Motion},  one would naively expect $\mu_1=\cdots=\mu_5=m_0$.  The mode integral returns a manifestly positive result, but the subsequent subtractions may render the net value of $P^\mathrm{ren}_{LR}$ negative. Because some of the subtraction terms explicitly contain time derivatives of the scale factor, renormalization does not simply involve removing the cutoff-dependent terms one would find in flat spacetime.  Note that we have separated the finite contributions of the matter fields from those that depend on the counterterms, although in some cases they are of  the same form. The subtraction term proportional to $(u_p \ddot{u}^*_p- \ddot{u}_pu_p^*)\mathcal{H}$ is unique in that way, since it does not depend on the cutoff, yet it is not renormalized by any of the counterterms. 

In order to derive equation (\ref{eq:PLR final}) we have not made any assumptions about the dynamics of gravity; the form of $P_{LR}$ essentially depends only on its couplings of matter.  If the mode functions $u_p$ of the graviton obey the equations of motion of general relativity, some of the expressions simplify. Say, in that case
\begin{equation}
	u_p \dot{u}^*_p-u_p^* \dot{u}_p=\frac{4i}{a^2 M_P^2}.
\end{equation}
More generally, for modes inside the horizon $u_p$ is given by equation (\ref{eq:h WKB}).

\subsection{Section Summary}
\label{sec:Section Summary}
In order to make sense of the divergent integrals that determine the power spectrum, we have introduced a set of Pauli-Villars massive regulator fields with Grassmann parity $\sigma_r$, whose masses  need to satisfy equation (\ref{eq:mi cond}) and (\ref{eq:mi add cond}). The introduction of these fields preserves diffeomorphism invariance and renders all  our mode integrals finite in the ultraviolet.
 
At momenta much smaller than the mass of the regulator fields, we  would expect the latter  to have no impact on the physical predictions of the theory, other than through the renormalization of the parameters of the low-energy theory. Indeed, when we attempt to decouple the regulators fields by sending their masses to infinity, we find  that we need to include divergent corrections to the action of the form (\ref{eq:counterterms}). The latter  are constrained by diffeomorphism invariance and can be sorted according to their mass dimension. The presence of an effective spacelike boundary in the spacetime at time $t$, where fields are not constrained to vanish, also forces us to include boundary counterterms like (\ref{eq:b counterterm}). This  construction then guarantees that the power spectrum remains finite both in the limit in which the cutoff is removed and the regulator masses are sent to infinity.

When the dust settles, the finite, renormalized value of the contributions to the power spectrum we shall need takes a relatively simple form, namely, that of equation (\ref{eq:PLR final}). Up to renormalization-dependent corrections, it is almost what one would get simply by imposing an ultraviolet cutoff on the divergent mode integral, and then subtracting the divergent cutoff-dependent pieces to render the integral finite \cite{Weinberg:2010wq}.
 
\section{Evaluation of the Energy Density}

A significant simplification in the evaluation of the different diagrams occurs because we are not directly interested in the power spectrum of the gravitational waves  (\ref{eq:Pdef}), but only in their effective energy density (\ref{eq:rho}) today. Suppose that preheating has concluded by some time $t_f$ in the early universe, and that we are interested in the density of the  produced gravitational waves at a much later time $t$.  Since the  relevant interactions  only occur before $t_f$, the power spectrum of gravitational waves at time $t\gg t_f$  follows from equations (\ref{eq:contributions}) simply by replacing  $t$  by $t_f$ in the upper limit of the time integrals. In order to evaluate the energy density of the produced gravitational waves at time $t$, we need to calculate a time average of the power spectrum over several oscillations of the gravitational wave. Because $P_{RR}$ is proportional to $u^2_p(t)\propto e^{-2i pt}/a^2$, and $P_{LL}$ is proportional to   $u_p^*{}^2(t)\propto e^{2i pt}/a^2$, these oscillatory contributions average out. In contrast, the contribution of $P_{LR}$ is proportional to $|u_p|^2\propto 1/a^2$ and hence survives the average. In  particular,  equation (\ref{eq:rho})    implies then that the spectral density indeed scales like radiation. Hence, all we really need to determine the spectral density  is the (renormalized) value of $P_{LR}$ in equation (\ref{eq:PLR final}).  

\subsection{The Preheating Stage}
\label{sec:The Reheating Stage}

At the end of inflation the inflaton oscillates around the minimum of its potential while, on average, the universe expands as if it were matter-dominated. The evolution of the  inflaton during that time is particularly simple in cosmic time $\tau$ \cite{Kofman:1997yn},
\begin{equation}
	a=a_0 \left(\frac{\tau}{\tau_0}\right)^{2/3}, \quad \bar{\phi}\approx \sqrt{\frac{8}{3}}M_P\frac{ \sin (m_\phi \tau)}{m_\phi\tau}.
\end{equation}
In what follows we shall set $M_\chi=0$ for simplicity. Since  in that case the effective mass of the matter field $\chi$ is $m_\chi^2=\lambda \bar{\phi}^2$ , it is  more convenient to solve  for the  time evolution of the mode functions $w_k$ in cosmic time too. Introducing the rescaled variable ${v_k\equiv w_k/a^{3/2}}$ and the dimensionless variables
\begin{equation}\label{eq:dimensionless variables}
 	x\equiv m_\phi \tau, \quad K\equiv \frac{k}{m_\phi}, 
	\quad P\equiv \frac{p}{m_\phi},
\end{equation}
 the former obeys
\begin{equation}
	\frac{d^2 v_K}{dx^2}+\left(\frac{4q_0\sin^2 x}{x^2}+\frac{K^2}{a^2}\right)v_K=0,  
\quad 
q_0\equiv \frac{2}{3}\frac{\lambda M_P^2}{m_\phi^2},
\end{equation}
where we have used that in a matter dominated universe $d^2 a/d\tau^2+aH^2/2=0$.  The mode equation can be cast as the Mathieu equation with a  time-dependent coefficient $q=q_0/x^2$. 

As the inflaton oscillates around the minimum of its potential, some of the matter modes experience parametric amplification.  This regime was analyzed in detail in reference \cite{Kofman:1997yn}.  According to this reference,  parametric resonance ends when  $q\approx 1/4$, that is, around 
\begin{equation}\label{eq:reheating ends}
	x_f\equiv  2\sqrt{q_0},
\end{equation}
and   particle production during preheating is efficient as long as the physical momenta obey the relation 
\begin{equation}\label{eq:resonant modes}
	\frac{K^2}{a^2}\leq  \frac{2}{\pi}\sqrt{q}. 
\end{equation}
Modes that undergo parametric resonance grow (on average) exponentially with cosmic time. The growth rate is very sensitive to the value of $q_0$ and the wave number \cite{Kofman:1997yn}, with a behavior that is hard to predict analytically.  Therefore,  gravitational wave production during preheating is typically studied using  numerical methods.

\subsection{Numerical Implementation}

 To determine the amount of gravitational waves produced by the modes that undergo parametric resonance  we shall  solve for  the  matter mode functions and perform the required integrals numerically.   The mode equations for $u_p$ and $w_k$, along with the  Einstein equations for the background, are solved numerically using the  \texttt{CVODE} routine in the Sundials suite \cite{hindmarsh2005sundials}.  The mode integrals are computed with the help of the \texttt{Cuhre} routine in the  \texttt{CUBA} integration library \cite{Hahn:2004fe}.
 
  Our numerical implementation naturally follows from our previous discussion, and was also  outlined in reference \cite{Weinberg:2010wq}.  It involves the evaluation of equation (\ref{eq:PLR final}) not in the limit $\Lambda\to \infty$, but    for a large arbitrary cutoff $\Lambda$. The error we commit by setting a cutoff at a finite value of $\Lambda$ can be estimated by looking up the terms in the integral that decay as $\Lambda\to\infty.$  For modes well inside the horizon, we expect the dominant error to scale as $1/\Lambda$ and contain no time derivatives of the background. On dimensional grounds alone, we thus expect the  relative error to be of  order
$
	\Delta P^\mathrm{ren}_{LR}/P^\mathrm{ren}_{LR}\sim p/\Lambda.
$

Following the discussion in subsection \ref{sec:The Reheating Stage}, we carry out the time integrals in cosmic time $\tau.$ To evaluate $P^\mathrm{ren}_{LR}$ in equation  (\ref{eq:PLR final}) we thus need to compute   integrals of the form
\begin{equation}\label{eq:cosmic time integral}
\int_{-\infty}^{\tau_f} d\tau_1 a(\tau_1) u_p(\tau_1) w_k(\tau_1) w_q(\tau_1)\,  e^{\epsilon \,\tau_1},
\end{equation}
where the factor $ e^{\epsilon\,\tau_1}$ amounts to the $i\epsilon$ prescription that we have kept implicit in previous formulas, and we assume that $\tau=-\infty$ denotes the asymptotic past. As it stands,   numerical evaluation of this integral is not feasible because it is  only practical  to set initial conditions at a finite time $\tau_i$, and because taking the limit $\epsilon\to 0$ numerically is too cumbersome. We split instead the integral as 
 \begin{equation}\label{eq:split integral}
\int_{-\infty}^{\tau_f} d\tau_1 f(\tau_1)\,  e^{\epsilon \tau_1}=
\int_{-\infty}^{\tau_i} d\tau_1 f(\tau_1)\,  e^{\epsilon \tau_1}+\int_{\tau_i}^{\tau_f} d\tau_1 f(\tau_1).
\end{equation}
As long as the evolution of the mode functions between $-\infty$ and $\tau_i$ remains adiabatic, we can approximate the mode functions by their adiabatic expansions  (\ref{eq:WKB expansions}) and thus  evaluate the first integral on the right analytically using integration by parts, as in equation (\ref{eq:integral expansion}). The $i\epsilon$ prescription implies that the contribution of the boundary at $\tau=-\infty$ vanishes, while the contribution of the boundary at $\tau_i$ can be readily evaluated to the desired adiabatic order.  We choose the initial scale factor and time $x_i\equiv m_\phi \tau_i$   to satisfy  
\begin{equation}
a_i\equiv 1, \quad x_i\equiv\frac{\pi}{2}, 
\end{equation}
 which roughly corresponds to the time at which the inflaton begins to oscillate around its minimum. For large values  of the $q_0$ parameter, the matter fields remain heavy throughout inflation, and the adiabatic approximation remains valid all the way past $\tau_i$ (deviations from adiabaticity occur when the matter fields effectively become massless \cite{Kofman:1997yn}.) The second integral on the right of equation (\ref{eq:split integral}) thus encompasses the reheating stage, and can be readily computed using numerical quadrature. In order to make sure that our results capture all of the preheating stage, we set the final time in equation (\ref{eq:cosmic time integral}) to that in equation (\ref{eq:reheating ends}), which we shall deem the ``end of preheating." Since we do not include backreaction,  reheating does not actually  result in a radiation-dominated  universe in our analysis, because the oscillating inflaton behaves as non-relativistic matter.    Backreaction  does play an important  during preheating at resonance parameters  $q_0\gtrsim10^3$ \cite{Kofman:1997yn}, which is why we mostly restrict our attention to
 \begin{equation}\label{eq:no backreaction}
 	q_0\lesssim 10^3.
 \end{equation}

\begin{figure}
\centering
\subfloat[]
{
 \includegraphics[width=7.5cm]{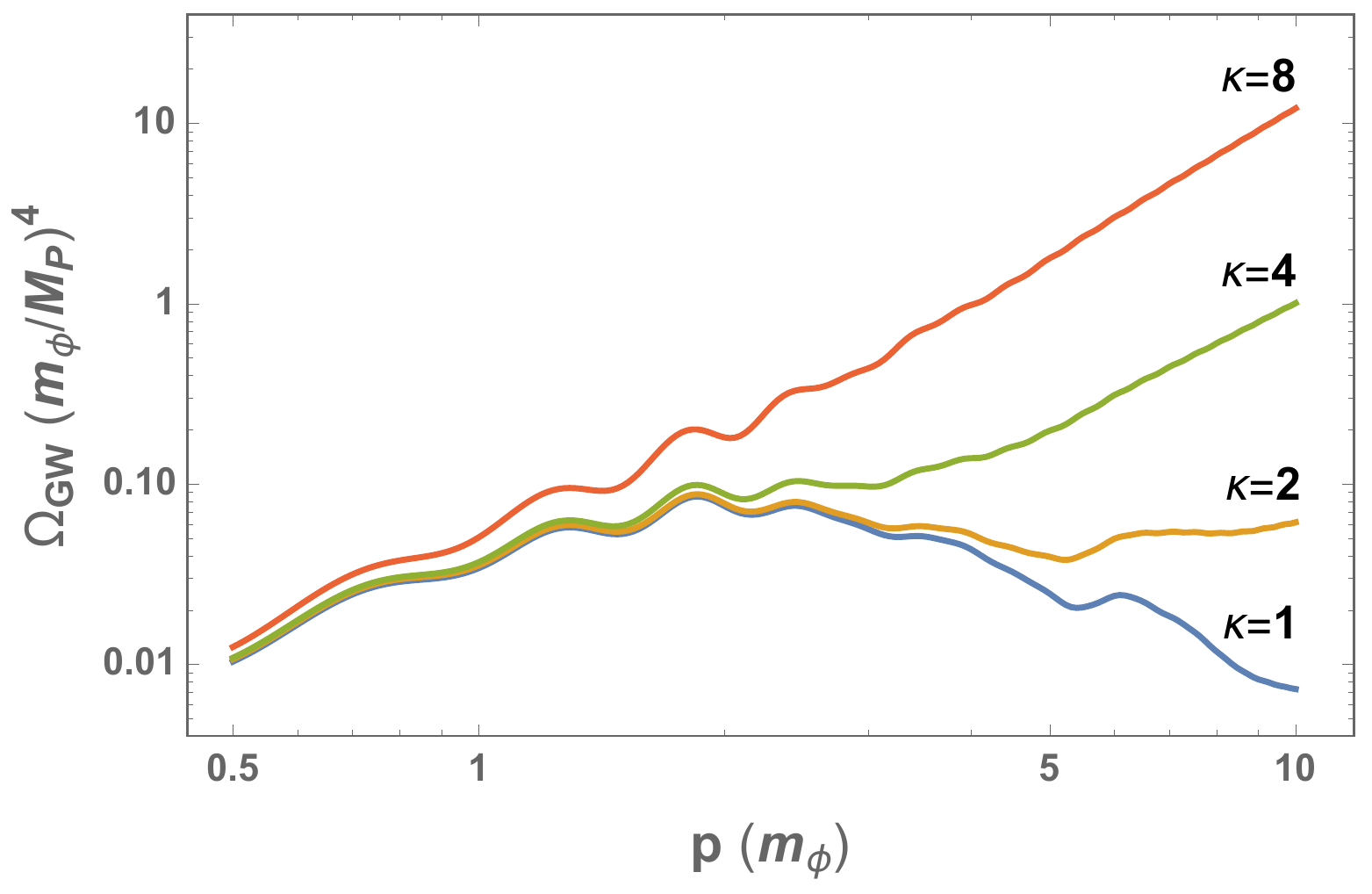}	
}
\subfloat[]
{

\includegraphics[width=7.5cm]{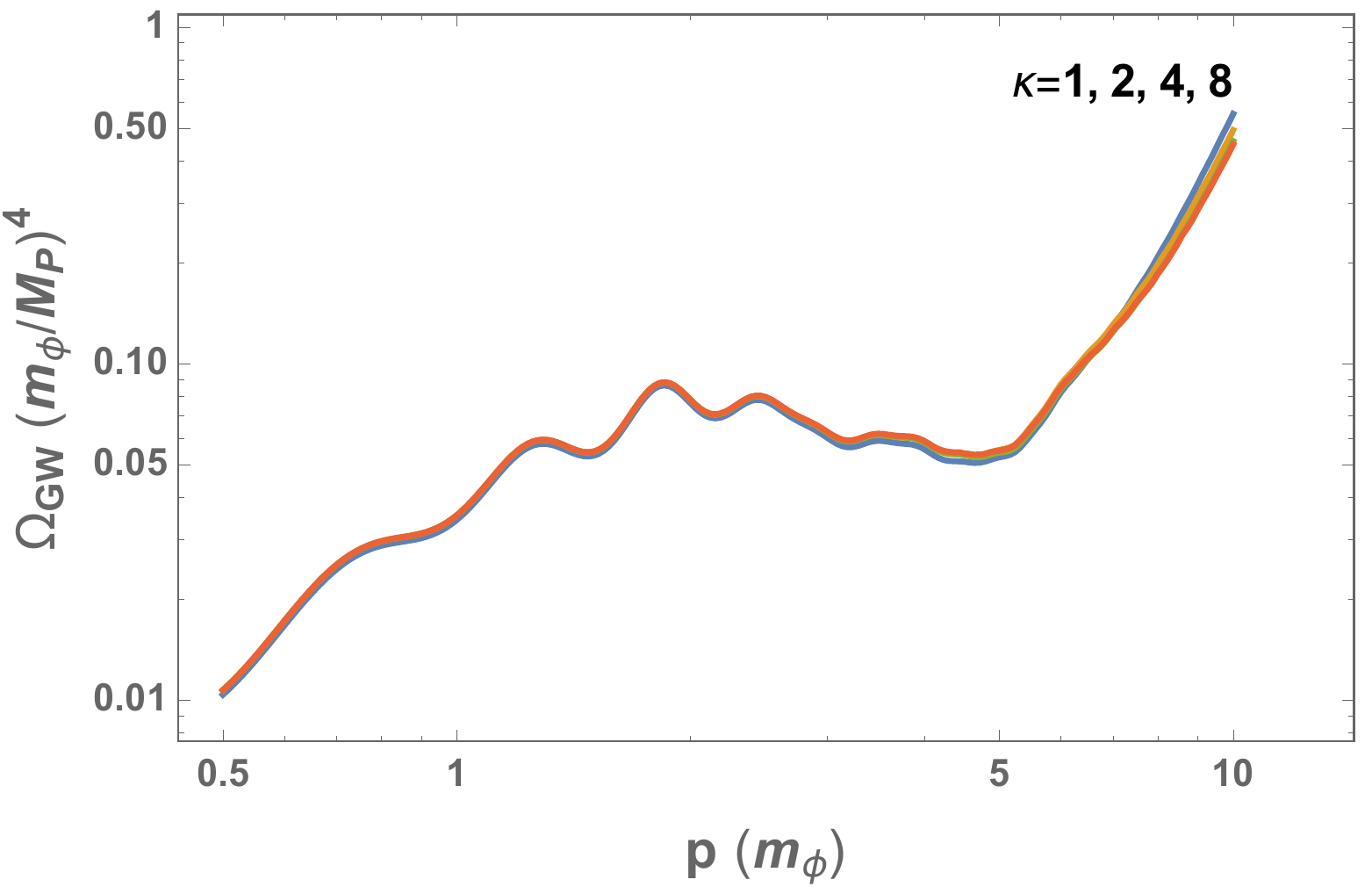}	
}
\caption{(a) Spectral density  in the $in$-$in$ formalism  for different momentum cutoffs (\ref{eq:Lambda}), with no subtraction terms included. In this case the  spectral density strongly depends on the cutoff. (b) Predicted spectral density in the $in$-$in$ formalism  with subtraction terms included.   This time the prediction is  clearly cutoff-independent. In both panels (a) and (b), $q_0=100$.\label{fig:ComparisonDependence}}
\end{figure}

The modes that undergo parametric resonance satisfy equation (\ref{eq:resonant modes}). Since these are the only modes for which we expect significant departures from adiabaticity, we shall thus choose a momentum cutoff $\Lambda$ at 
\begin{equation}\label{eq:Lambda}
	\frac{\Lambda}{m_\phi}=2 \kappa\times \cdot\left(\frac{2}{\pi} \frac{\sqrt{q_0}}{x_f}\right)^{1/2}\,a(x_f),
\end{equation}
where $\kappa$ is a  parameter that controls the size of the cutoff. Our default choice is $\kappa=1$.  Note that $a^2(x)/x$ is an increasing function of $x$ during preheating, so  such a  cutoff ensures that modes with $K>\Lambda$  never satisfy the condition for effective resonance.  By changing the value of $\kappa$ we can estimate the size of the errors associated with  the finite cutoff.  Some of the  oscillatory integrands lead to slow convergence, and to improve the speed of the calculations we  prescribe a relative precision of $10^{-3}$.  This ought to be sufficient at large values of $q_0$, but may yield large relative errors in the final spectral density when integrals and subtraction terms cancel to one part in a thousand. Figure \ref{fig:ComparisonDependence} shows for  example  how the predicted spectral density depends on the value of $\kappa$ at $q_0=100$, and how the latter is affected by the cutoff-dependent subtraction terms in equation (\ref{eq:PLR final}). 
 Because  the ratio of the leading subtraction terms in equation (\ref{eq:PLR}) to those that depend on the renormalization conditions is of order $(p/\Lambda)^3$, we do not expect the latter to have much of an impact on the predictions of the spectral density when parametric resonance is effective.

Finally, one should  bear in mind that the net density of gravitational waves  also contains the contribution from the  free-field fluctuations  in  equation (\ref{eq:vev}),
\begin{equation}\label{eq:omega tree}
	\Omega^\mathrm{tree}_{GW}\approx \frac{3\left(\pi^2/2\right)^{1/3}}{128} \frac{1}{q_0^{1/3}}\left(\frac{p}{a_i m_\phi}\right)^4 \frac{m_\phi^2}{M_P^2}.
\end{equation}
Because this contribution is proportional to $(m_\phi/M_P)^2$,  as opposed to   the $(m_\phi/M_P)^4$ proportionality of the  one-loop corrections,  it typically dominates  at sufficiently small values of $q_0$, unless $m_\phi$ is  close to Planckian.  Note that the tree-level density (\ref{eq:omega tree}) depends on $q_0$ because the former is evaluated at the end of reheating, equation (\ref{eq:reheating ends}), which does depend on that parameter. 

\subsection{Results}

\begin{figure}
 	\includegraphics{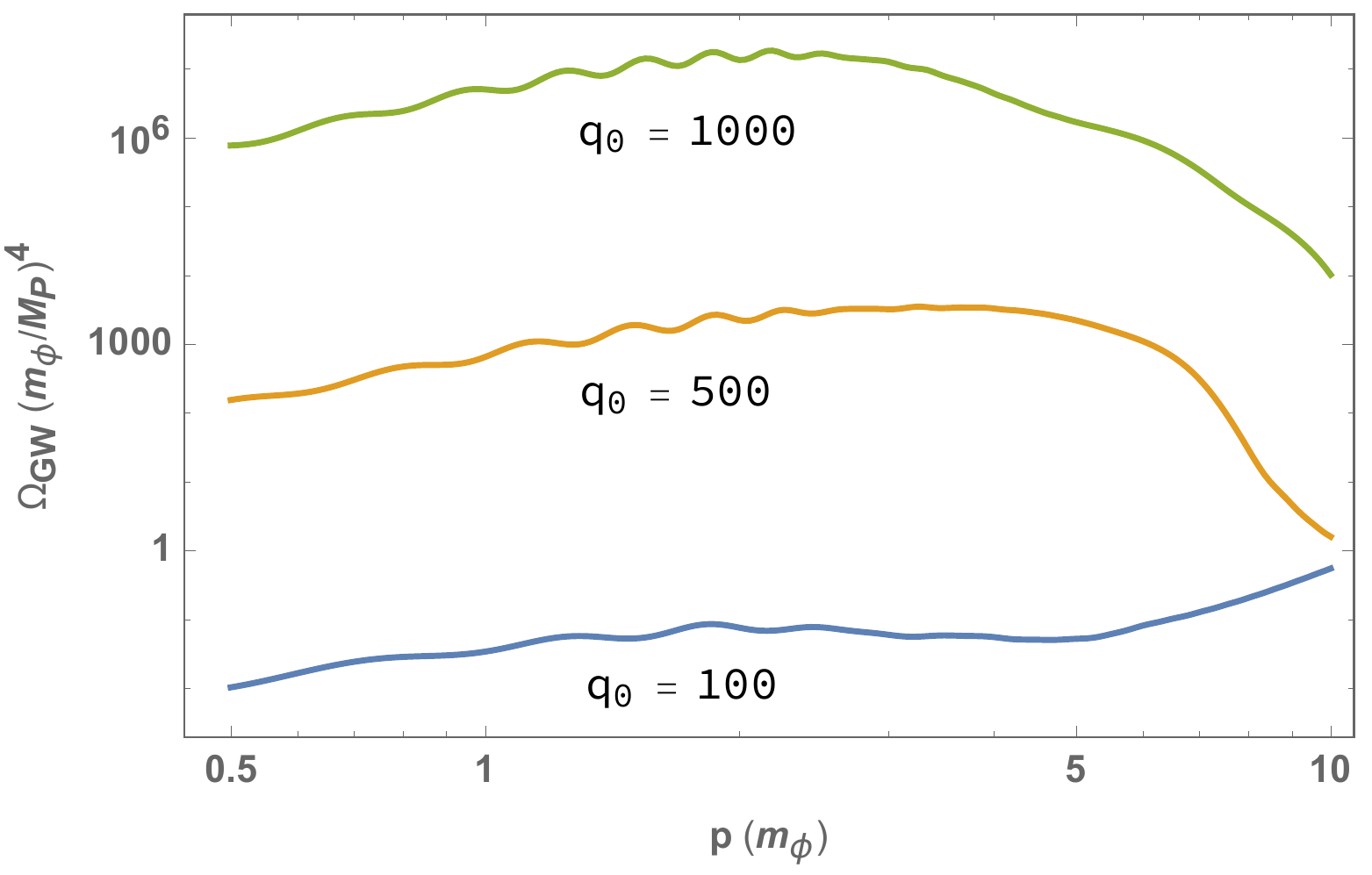}
	\caption{Spectral density of the produced gravitational waves for different values of the resonance parameter $q_0$. All density parameters are evaluated at the end of preheating, equation (\ref{eq:reheating ends}). We set $a_i=1$.
\label{fig:density parameters}} 
\end{figure}

Our main numerical results are summarized in figure \ref{fig:density parameters}, which shows the predicted spectral densities  in the $in$-$in$ formalism for different resonance parameters $q_0$. As seen on the figure, the   gravitational wave signal strongly depends on  $q_0$.  In fact, the behavior of the  mode functions during parametric resonance suggests that this dependence is exponential. Such an exponential growth, however, cannot continue at arbitrary large values of $q_0$. As parametric resonance becomes increasingly effective,  backreaction on the inflaton oscillations quenches the effects of parametric resonance \cite{Kofman:1997yn, Figueroa:2017vfa}. The spectral density is quoted  in units of $(m_\phi/M_P)^4$ and is thus quite sensitive to the mass of the scalar $\phi$.  Note, in particular, that the strength of the gravitational waves essentially depends on just  dimensionless parameters, $q_0$ and $m_\phi/M_P$ (recall that we have set $M_\chi\equiv M_0=0$ for simplicity.)

In order to compute the density parameter and the physical frequency of the waves \emph{today}, $f_0$, we need to follow the evolution of the scale factor and the energy density. Using standard results we find
\begin{equation}
	f_0\equiv \frac{1}{2\pi}\frac{p}{a_0}\approx 6.8\cdot  10^{10} \left(\frac{g_{*S}^0}{g_{*S}^\mathrm{rh}}\right)^{1/3}\left(\frac{g_{*}^\mathrm{rh}}{g_{*}^0}\right)^{1/4}\left(\frac{a_\mathrm{i}}{a_\mathrm{rh}}\right)^{1/4} \left(\frac{m_\phi}{M_P}\right)^{1/2} \frac{p}{a_i m_\phi} \,\mathrm{Hz},
\end{equation}
where the index ``$i$" refers to the beginning of reheating,  ``$\mathrm{rh}$" to its end, ``0" to today,  and the different factors $g$ are those in reference \cite{Kolb:1990vq}. The density parameter today differs from that at the time of reheating by about  four orders of magnitude, 
\begin{equation}\label{eq:omega today}
	\Omega_{GW}^0= 
	\frac{g_*^\mathrm{rh}}{g_*^0}\left(\frac{g_{*S}^0}{g_{*S}^\mathrm{rh}}\right)^{4/3}
	{\Omega^0_{\mathrm{rad}}}\, \Omega_{GW}, 
	\quad \text{with} \quad
	\Omega^0_\mathrm{rad}\approx 9.2 \cdot 10^{-5}.
\end{equation}
It is important to note that, in the absence of backreaction, the inflaton energy density redshifts  as $1/a^3$ during reheating, while the energy density of the gravitational waves scales as $1/a^4$. Hence, the value of $\Omega_{GW}$ is somewhat sensitive to the time we designate as the end of reheating. 

Unfortunately an inflationary model with a  purely quadratic potential is strongly disfavored by a combination of BICEP2/Keck Array and Planck collaboration data  \cite{Akrami:2018odb}. For illustration, we shall frame these results, instead, with the example of  an  inflationary model with scalar  potential
\begin{equation}\label{eq:Starobinsky V}
	V(\phi)=\frac{3}{4}M_P^2 m_\phi^2 \left[1-\exp\left(-\sqrt{2/3} \, \phi/M_P\right)\right]^2,
\end{equation}
 which  happens to be among those that best fit  current observations  \cite{Akrami:2018odb}.  
The   potential (\ref{eq:Starobinsky V}) is that  of the  Starobinsky and Higgs inflationary  models \cite{Martin:2013tda}, although the couplings to matter  that we assume are not necessarily the ones  in   those models. There is a wide variety of scenarios  with a potential that can be approximated by  (\ref{eq:Starobinsky V}) during inflation, at $\phi\gg M_P$,  although they may not agree with it globally \cite{Martin:2013tda}. 

The  potential (\ref{eq:Starobinsky V}) is quadratic  around $\phi=0$, with relative deviations from pure quadratic behavior that remain smaller than about  $50\%$ at $|\phi|\lesssim M_P/2$. Therefore, a quadratic potential ought to be a good approximation after the first few oscillations of the inflaton.    In the model (\ref{eq:Starobinsky V}) the value of $m_\phi$ is determined by the observed amplitude of the primordial scalar perturbations,  $m_\phi/M_P\approx 1.2\cdot 10^{-5}$ \cite{Martin:2013tda}. Therefore, in this case the signal would peak at present frequencies of about $2\times10^8$ Hz, which is  three orders of magnitude above  the highest  frequencies probed by current and near-future detectors \cite{Moore:2014lga}.  

The value of the coupling $\lambda$, and hence $q_0$, is poorly constrained.  We shall only demand that the  induced radiative corrections to the potential (in flat spacetime) remain subdominant both  during inflation and  the oscillation phase thereafter, 
\begin{equation}
 \frac{\lambda^2}{32\pi^2}\phi^4 \log \frac{\phi}{\mu}\ll \frac{3}{4}M_P^2 m_\phi^2
 \quad
 \text{and}
 \quad
 \frac{\lambda^2}{32\pi^2}\phi^4 \log \frac{\phi}{\mu}\ll \frac{1}{2}m_\phi^2 \phi^2.
\end{equation}
Assuming that the $\log$ is of order one, and using that  $\phi\lesssim 6M_P$  during inflation, we find  $q_0$ cannot be higher than about $10^5$ if we are to trust the tree-level potential. Yet we cannot explore this full parameter range because backreaction is only negligible when $q_0\lesssim 1000$. In that interval,  the curves in figure \ref{fig:density parameters} indicate that the gravitational wave density would be too weak to be detectable  by  near future detectors \cite{Moore:2014lga} even if the signal happened to fall in the appropriate frequency range. In order to explore the parameter space of detectable signals, we would need to add backreaction to our calculations. 

\subsection{Comparison with Previous Approaches}
 
As we have emphasized in the introduction,  previous estimates of gravitational wave production during reheating \cite{Dufaux:2007pt}  do not proceed from equation (\ref{eq:vev}).  Numerical estimates  rely instead  on an ensemble of classical universes in which the  relation between matter fields and the sourced gravitational waves  is the same as that in  equation (\ref{eq:semiclassical}),
\begin{equation}
\bar{h}(t,\vec{p}) \bar{h}(t,-\vec{p})=
	\int^{t} \! dt_1 \!  \int_{t_i} ^{t} \!  dt_2 \, D_{ p}^R(t;t_1)  D_{ p}^R(t;t_2) 
	\bar{S}_1(t_1,\vec{p}) \bar{S}_1(t_2,-\vec{p}).
\end{equation}
In this equation the bars denote classical fields, and the $\bar{S}_1$ are  obtained from equation (\ref{eq:S1})   simply by replacing the matter fields $\chi_{\vec{k}}$ by  numerically evaluated solutions $\bar\chi_{\vec{k}}$ of the classical field equations.   The necessary initial conditions are chosen from a random distribution that mimics the statistical properties of the matter fields in the vacuum, be it after the end of inflation, or well within inflation. Let $\bar{\chi}_{\vec{k}}$ be  such a solution.  Because the mode functions $w_k$ and their complex conjugates $w_k^*$ are linearly independent solutions of the field equations, we can expand the  $\bar{\chi}_{\vec{k}}$ as
\begin{equation}
	\bar{\chi}_{\vec{k}}(t)=\alpha_{\vec{k}} \, w_k(t)+\alpha_{-\vec{k}}^* \,w_k^*(t),
\end{equation}
 where  we have enforced the reality of the fields, and the $\alpha_{\vec{k}}$ are constant random coefficients drawn from an appropriate Gaussian distribution. The latter is determined by the requirement that it reproduce the statistical properties of the  fields in the quantum vacuum, namely,
 \begin{equation}\label{eq:alpha moments}
[\bar{\chi}_{\vec{k}}]=\langle \chi_{\vec{k}}\rangle=0 ,
\quad
[\bar{\chi}_{\vec{k}}\bar{\chi}_{\vec{q}}]=\langle \chi_{\vec{k}} \chi_{\vec{q}}\rangle=|w_{k}|^2 \delta_{\vec{k}+\vec{q}},
 \end{equation}
where  $\langle \cdots\rangle$ denotes vacuum expectation value, and $[ \cdots]$ expectation in the random distribution used in the numerical simulations.  If we regard the $\alpha_{\vec{k}}$ and their complex conjugates $\alpha_{\vec{k}}^*$ as independent variables, equations (\ref{eq:alpha moments}) imply 
\begin{equation}
[\alpha_{\vec{k}}]=[\alpha_{\vec{k}}^*]=0,
\quad
 [\alpha_{\vec{k}}\,  \alpha_{\vec{q}}]= [\alpha_{\vec{k}}^* \,\alpha_{\vec{q}}^*]=0,
 \quad \textrm{and} \quad
 [\alpha_{\vec{k}} \,\alpha_{\vec{q}}^*]\equiv [\alpha_{\vec{q}}^*\,\alpha_{\vec{k}} ]=\frac{\delta_{\vec{k},{\vec{q}}}}{2},
\end{equation}
which,  in a way, mimics the behavior of creation/annihilation operators.

An estimate of the power spectrum of gravitational waves may be  obtained by decomposing the momentum vector $\vec{p}$ into its magnitude $p$ and direction $\hat{p}$,  and averaging over the angular variables $\hat{p}$  \cite{Dufaux:2007pt}. For our purposes it is more convenient to invoke  ergodicity  and replace the angular average by  its expectation in an ensemble of simulations,
\begin{equation}\label{eq:classical P}
	[\bar{h}(t,\vec{p}) \bar{h}(t,-\vec{p})]=
	\int_{t_i}^{t} \! dt_1 \int_{t_i}^{t} \! dt_2 \, D^R_{p}(t;t_1)  D^R_{p}(t;t_2) [ \bar{S}_1(t_1,\vec{p}) \bar{S}_1(t_2,-\vec{p})],
\end{equation}
 where   the correlation of the classical sources in the simulation obeys
 \begin{equation}\label{eq:classical SS}
 [ \bar{S}(t_1,\vec{p}) \bar{S}(t_2,-\vec{p})]_\mathrm{conn}=
	\frac{a^2(t_1) a^2(t_2)}{8(2\pi)^3}\int d^3k\, 
	\mathrm{Re}[G_k(t_1;t_2)]\mathrm{Re}[G_q(t_1;t_2)]
	k^4 \sin^4\theta.
 \end{equation}
We have restricted our attention to the connected piece of the correlation, because  translational invariance of the random distribution again implies,  on  average, that the disconnected piece only sources ``gravitational waves" of zero momentum. 

In order to compare previous numerical  estimates with the predictions of the $in$-$in$ formalism, we shall cast  equation (\ref{eq:classical SS}) also as a sum of squares. Dropping terms that oscillate at time $t$ we find 
\begin{equation}\label{eq:final power rdm}
\begin{split}
[\bar{h}(t,\vec{p}) \bar{h}(t,-\vec{p})]^\mathrm{conn}_\mathrm{rdm}=
\frac{|u_p(t)|^2}{32(2\pi)^3}\int d^3 k\, k^4 \sin^4\theta  \Bigg\{ \left| \int_{t_i}^{t_f} dt_1 a^2(t_1) u_p(t_1) w_k(t_1) w_q(t_1) \right|^2 \\
+\left|\int_{t_i}^{t} dt_1 a^2(t_1) u^*_p(t_1) w_k(t_1) w_q(t_1) \right|^2 +\cdots \Bigg\},
\end{split}
\end{equation}
where the dots stand for the   integrals that arise from the six remaining ways of conjugating or not conjugating each mode function individually  (since there are three mode functions in the integrand, there is a total of $2^3$ different combinations.) Numerical estimates do not actually calculate the expression (\ref{eq:final power rdm}) directly. Instead, they solve for the matter fields by evolving a discretized universe and use these classical fields as  the sources of gravitational waves \cite{Felder:2000hq,Frolov:2008hy,Easther:2010qz,Huang:2011gf}. In any case, equation (\ref{eq:final power rdm})  is constructed to reproduce what these codes are aiming to compute. At this point it is inconsequential  whether  the backreaction  is taken into account. The latter affects the actual values of the mode functions $w_k$ and the scale factor, but not the actual  form of equation (\ref{eq:final power rdm}).

There are also some analytical estimates of gravitational wave production \cite{Figueroa:2017vfa}, in addition to the original estimate of reference \cite{Khlebnikov:1997di}. The former begin with  equation (\ref{eq:classical P}), with the expectation in the ensemble of simulations replaced by the vacuum expectation value. Such a substitution does not quite return our unrenormalized expression for $P_{LR}$,  because the  Green's functions for the gravitational waves differ. It is also worth pointing out that these analytical estimates and their numerical simulation counterparts are not calculating the same quantity. The sources are quadratic in the matter fields,  so  even if equations (\ref{eq:alpha moments}) hold, 
$\langle S_1(t_1,\vec{p}) S_1(t_2,-\vec{p})\rangle$ differs in general from  $[ \bar{S}_1(t,1,\vec{p}) \bar{S}_2(t_2,-\vec{p})]$. In any case, such analytical estimates do seem to agree with   the numerical simulations, which are the ones we shall focus on.

The  first difference between equations (\ref{eq:final power rdm}) and (\ref{eq:PLR final}) lies in the lower limits of integration. In the $in$-$in$ formalism the initial time is set in the asymptotic past, at $\tau_i=-\infty$, where  the $i\epsilon$ prescription eliminates the dependence of the integral on the fields at the past boundary. If we insist in carrying out the integral from  a finite lower  boundary  $t_i$, we should  add the missing portion of the integral, as in equation  (\ref{eq:split integral}). This missing piece can be estimated analytically as long as the evolution of the modes is adiabatic, and only involves the values of the mode functions at $t_i$.  For sufficiently  small values of the resonance parameter $q_0$, the magnitude of the boundary term at $t_i$ can be  comparable to that of the integral between $t_i$ to $t_f$. If these boundary terms are not taken into account, the correlation function ends up depending on the mode functions at the initial time $t_i$,  and contains additional oscillatory terms that would not be present otherwise.   In particular, because the  dependence on the past boundary persists in the ultraviolet,  the divergent piece of the mode integral (\ref{eq:final power rdm}) ends up also depending on the initial time $t_i$ and not just on a local expression defined at time $t_f$.
 
Leaving the previous differences aside, and focusing just on the integrals in both (\ref{eq:final power rdm}) and (\ref{eq:PLR final}), one may observe  that  both expressions would  agree if the integrals were insensitive to the phase of the mode functions.  This is to some extent what happens when there is strong particle production, as during preheating. In general, we can cast the matter mode functions in the form
\begin{equation}
	w_k=\frac{1}{a\sqrt{2\omega_k}}\left[\alpha_k(t)\exp\left(-i\int^t \omega _k\,  dt_1\right)+\beta_k(t)\exp\left(i\int^t \omega _k \,dt_1\right)\right],
\end{equation}
with Bogolubov coefficients $\alpha_k$ and $\beta_k$ that are constrained to satisfy ${|\alpha_k|^2-|\beta_k^2|=1}$. The matter mode functions are those of  the $in$ vacuum, that is  $\alpha_k\to1, \beta_k\to0$ in the asymptotic past, but  during preheating  $|\beta_k|^2$ grows to large values for those modes that experience parametric resonance. In this limit, up to a phase,  $\alpha_k\sim \beta_k$, and, therefore $w_k\sim w_k^*$. In that sense, the expectation that a classical analysis is justified in the presence of strong particle production bears out. Note, however, that the same argument does not apply to the tensor mode functions $u_p$, since the evolution of the latter remains adiabatic throughout. 

\begin{figure}
\begin{center}
\subfloat[$q_0=1000$, $p=2m_\phi$]
{
\includegraphics[width=7.5cm]{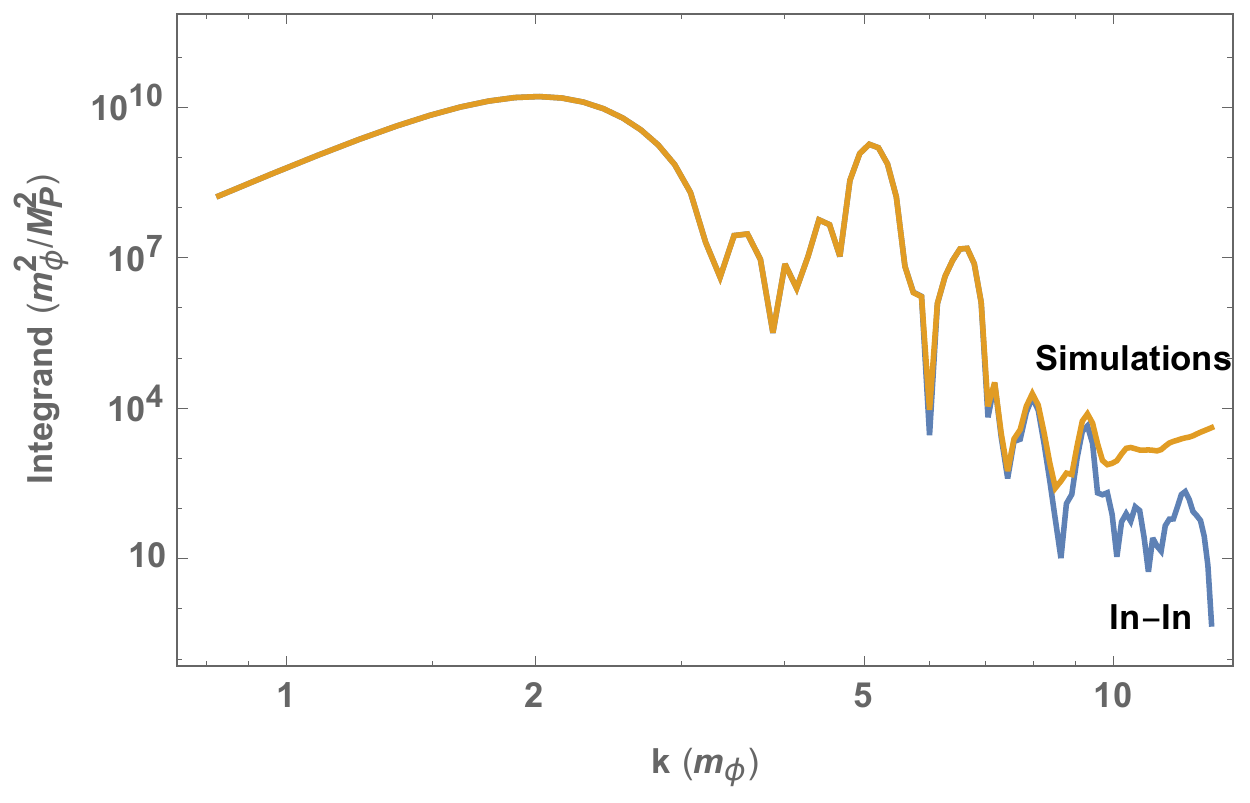}
}
\subfloat[$q_0=100$, $p=2m_\phi$]
{
\includegraphics[width=7.5cm]{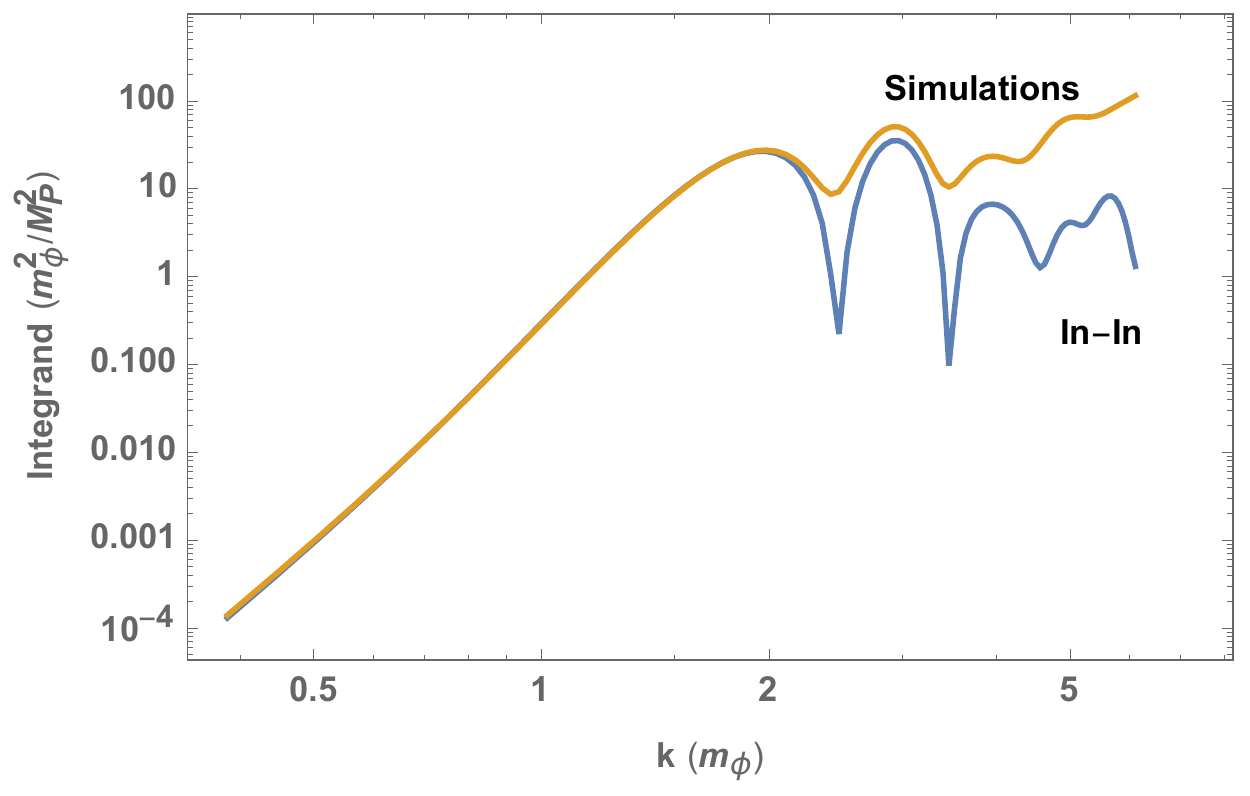}
}
\caption{Comparison of momentum integrands (per logarithmic interval $dk/k$) in the $in$-$in$ formalism and as expected from numerical simulations. No backreaction is included in either case. \label{fig:integrands}} 
\end{center}
\end{figure}

The difference between the two approaches is apparent in figure \ref{fig:ComparisonDependence} (b), which shows how the predictions from the $in$-$in$ formalism and the numerical simulations significantly diverge at large values of $p$.  To gain a quantitative understanding, we  also compare the numerically computed momentum integrands (per logarithmic $k$)  in equations (\ref{eq:PLR final}) and equation (\ref{eq:final power rdm}) for $\sin \theta=1$ and various values of $q_0$ in figure \ref{fig:integrands}.  At  momenta $k$ around the main peaks  of the integrand, the  difference between the integrands is small at large $q_0$, but significant  at small values of  $q_0$, in agreement with the expectation that the difference ought to be small when particle production is effective. 

Yet perhaps the most important difference between equations  (\ref{eq:PLR final}) and  (\ref{eq:final power rdm}) concerns the subtraction terms that  the former contains and the latter lacks.  Because the mode integral (\ref{eq:final power rdm}) diverges in the ultraviolet, no matter how effective parametric resonance is, the dominant contribution to the integral (\ref{eq:final power rdm}) stems from the modes around the cutoff $\Lambda$, provided the latter is sufficiently large. This has been noted before, say, in reference \cite{Figueroa:2017vfa}, which also discusses in passing how numerical simulations deal with this problem. Our results can be used in fact to justify the approach followed by the simulations. Within the  $in$-$in$ formalism, at large $q_0$,  the subtraction of the cutoff-dependent terms in  (\ref{eq:PLR final}) does not  have much of an impact on the final answer, as long as the integration range in momenta is restricted only  to the modes that experience  parametric resonance. If, on the other hand, the range of momenta is extended far beyond the latter, the subtraction of these terms simply cancels the contribution of such ``ultraviolet" modes, which   do not experience parametric resonance by definition.  In that sense, it is numerically more efficient to restrict the mode integral to  those modes that experience significant deviations from adiabaticity.

Unfortunately,   the existing analyses that rely on  actual numerical simulations, such as \cite{Easther:2006gt, Figueroa:2017vfa},  have concentrated in large values of the resonance parameter $q_0$, for which the backreaction of the matter fields on the evolution of the inflaton  is expected to be significant. Therefore, it is not possible to  compare  their predictions directly  with ours. Nevertheless, the shapes of the corresponding spectral densities appear to be in rough qualitative agreement.

\section{Summary and Conclusions}

In this manuscript we have developed a framework to predict the energy density of the gravitational waves produced during reheating from first principles. Our estimate is grounded on the $in$-$in$ formalism, which we employ to directly compute the expected energy density of the gravitational waves produced during  that stage. It involves the leading terms in an expansion of the expected density  to lowest order in inverse  powers of the Planck mass, which happens to arise from a  Feynman diagram in which the scalar field the inflaton decays into runs in  a loop. The actual prediction essentially requires the calculation of a mode  integral over the mode functions of the matter fields and the gravitational waves, which is arguably  simpler than the traditional approaches that rely  on evolving a discretized universe. 

In order to obtain a well-defined prediction out of the divergent mode integrals, we had to pay particular attention to the regularization and renormalization of general relativity coupled to a  scalar. Because  the mode functions of the different fields during preheating  are not analytically known, and  diffeomorphism invariance is crucial to the renormalizability of the theory, we have opted for an implementation of Pauli-Villars  regularization that involves the introduction of Grassmann-odd scalar fields similar to the Faddeev-Popov ghosts of gauge theories.  New divergences  appear as we attempt to decouple these ghosts, but the latter can be absorbed into appropriate counterterms in the action. The required counterterms are those encountered in  the classic {$in$-$out$} analyses, but, in addition, they need to be supplemented with  counterterms in a boundary action defined on the spacelike  surface  at the final time $t$. At lowest order in derivatives, the latter is proportional to the York-Hawking-Gibbons boundary action.  

At large values of the resonance parameter, $q_0\gg 100$, in the absence of backreaction, our predictions seem to agree with those expected from numerical simulations that evolve a classical universe from appropriate initial conditions.  At $q_0\ll 100$, however, these simulations cease to be applicable, and our approach becomes the only way to obtain accurate estimates. In that sense, some of our results can be taken to be the first accurate predictions of gravitational wave production during preheating  at small values of the resonance parameter $q_0$.  Since we can expect a sufficiently strong signal of gravitational waves after inflation when parametric resonance is very effective, the question may appear academic. But since our analysis is quite model-independent, it ought to apply to scenarios  in which the resonance parameter remains moderate, and the traditional numerical computations are not justified.

The main drawback of our study thus far is the absence of backreaction on the evolution of the background inflaton and metric. One should be able to include the latter using  exactly the same formalism we have presented here,    along the  lines  discussed in appendix \ref{sec:Effective Equations of Motion}. Leaving renormalization aside, it is plausible that the agreement between numerical simulations and the $in$-$in$ formalism will also persist once backreaction is included, although in our opinion the issue deserves a more rigorous justification that the conventional argument invoking particle production and large occupation numbers. 

More generally, we believe that our analysis has shed further light into renormalization in the $in$-$in$ formalism, a topic that has not received much attention in the literature, and appears to have a richer structure than its $in$-$out$ counterpart, particularly when one is interested in expectation values at finite times. In that sense, some of our method should be applicable to the calculation of  correlation functions in a wide array of  cosmological scenarios.

\appendix
\section{Polarization Tensors}
\label{sec:Polarization Tensors}
As mentioned in the main text, it is useful to decompose the transverse and traceless part of the metric perturbations into a set of decoupled modes of definite helicity, as in equation (\ref{eq:h expansion}). For $\sigma=\pm 2$, the tensors $Q_{ij}{}^\sigma$ are defined by
\begin{equation}
	Q_{ij}{}^{(+2)}(\vec{p})=\hat{\epsilon}^+_i (\vec{p})\hat{\epsilon}^+_j(\vec{p}),
	\quad
	Q_{ij}{}^{(-2)}(\vec{p})=\hat{\epsilon}^-_i (\vec{p})\hat{\epsilon}^-_j(\vec{p}),
\end{equation} 
where the  vectors $\epsilon^\pm$ are two complex orthonormal transverse vectors satisfying
\begin{equation}
\hat{\epsilon}^{\sigma_1}(\vec{p})^* \cdot \hat{\epsilon}^{\sigma_2}(\vec{p})=\delta^{\sigma_1\sigma_2}, 
\quad
\vec{p}\cdot \hat{\epsilon}^\pm(\vec{p})=0\quad \textrm{and}\quad\vec{p}\times  \hat{\epsilon}^\pm(\vec{p})=\mp ip \hat{\epsilon}^\pm(\vec{p}). 
\end{equation}
When $\vec{p}$ points along the $z$ direction, these vectors can be taken to be 
\begin{equation}
	\hat{\epsilon}^\pm(\hat{z})=\frac{1}{\sqrt{2}} \left(\hat{e}_x\pm  i\hat{e}_y\right),
	\quad
	\hat{\epsilon}^\pm(-\hat{z})=-\hat{\epsilon}^\mp(\hat{z}),
\end{equation}
where $\hat{e}_x$ and $\hat{e}_y$ respectively are unit vectors along the $x$ and $y$ directions. If $\vec{p}$ does not point along the $z$ direction, $\epsilon^\pm(\vec{p})$ is obtained by a standard proper rotation of the latter.  In particular, note that $\hat{\epsilon}^\sigma  (\vec{p})^*=\hat{\epsilon}^{-\sigma}(\vec{p})$, which implies that $Q_{ij}{}^\sigma$ is traceless, and ${\hat{\epsilon}^\sigma(-\vec{p})=- \hat{\epsilon}^{-\sigma}(\vec{p})}$, which implies that $Q_{ij}{}^{\sigma_1} (\vec{p})Q^{ij}{}^{\sigma_2} (-\vec{p})=\delta^{\sigma_1\sigma_2}$.

\section{Boundary Terms in the Interaction}
\label{sec:Boundary Terms in the Interaction}

To see under what conditions  boundary terms contribute to the expectation of an observable,  consider an interaction Hamiltonian that contains a total derivative,
\begin{equation}
	H_I(t)=\frac{dB_I}{dt}+\bar{H}_I.
\end{equation}
 In that case, the time evolution operator in the interaction picture is
\begin{equation}
U_I= e^{-iB_I(t)}T\exp\left(-i\int^t \bar{H}_I(t_1)dt_1\right),
\end{equation}
where $T$ denotes time ordering, which indeed satisfies ${idU_I/dt=H_I(t) U_I(t)}$. The expectation value of an observable $\mathcal{O}$ in the $in$-$in$ formalism  is then
\begin{equation}
	\langle \mathcal{O}(t)\rangle=\left\langle \bar{T} \exp\left(i\int^t \bar H_I(t_1)dt_1\right)e^{iB_I(t)}\mathcal{O}_I(t)e^{-iB_I(t)}T \exp\left(-i\int^t \bar H_I(t_1)dt_1\right)\right\rangle.
\end{equation}
Therefore, if  $[\mathcal{O}_I(t),B_I(T)]\neq0$  the boundary term does    contribute to the expectation. This is typically what happens when either the interaction or the operator $\mathcal{O}_I$ depend on the canonical momenta.  In the main text we had to calculate the expectation of a function of $h$, with an interaction that contained its time derivatives (which are proportional to the canonical momenta.) This is why we should expect the expectation to depend on some of the  boundary terms in the action.

\section{Effective Equations of Motion}
\label{sec:Effective Equations of Motion}

Our estimate of the energy density of gravitational waves can be circumscribed in a framework that aims to derive quantum corrections to the classical gravitational equations. To arrive at these quantum-corrected equations we shall begin with the quantum effective action in the $in$-$in$ formalism, $\Gamma$, which is a functional of a set of field expectations $\bar{g}_{\mu\nu}^L$, $\bar{g}_{\mu\nu}^R$, $\bar{\phi}^L$, $\bar{\phi}^R$,  $\bar{\chi}^L$, $\bar{\chi}^R$  \cite{Calzetta:1986ey}. It is quite useful to regard the latter as fields defined on the two different branches of  the  Schwinger-Keldysh contour: The $R$ branch runs from $-\infty$ to $t_B$, and the $L$ branch from $t_B$ back to $-\infty$ \cite{Berges:2004yj}. In this picture, the action of the theory is a functional of a single set of fields, and we can borrow    all the results on the quantum effective action from the conventional  $in$-$out$ formalism essentially without modification.

In this context, then, just as in the $in$-$out$ formalism, the quantum corrected Einstein equations of motion follow from the condition
$
	\delta \Gamma/\delta \bar{g}_{\mu\nu}(t,\vec{x})=0,
$
which translates into the path integral equation
\begin{equation}
	\left\langle \frac{\delta S}{\delta g_{\mu\nu}}\right\rangle_{1PI}\equiv	\int\limits_{1PI} D \delta g_{\mu\nu} D\delta\phi D\delta \chi  \, \frac{\delta S}{\delta g_{\mu\nu}(t,\vec{x})} \exp\left(iS[\bar g+\delta g,\bar\phi+\delta\phi, \bar\chi+\delta\chi]\right)=0.
\end{equation}
Here we have used that the effective action can be written as the sum of all connected one-particle-irreducible diagrams  in a theory in which the fields are the sum of a background value plus quantum fluctuations one integrates over \cite{Weinberg:1996kr}. Because the path integral runs over fields defined on the Schwinger-Keldysh contour, the insertion of $\delta S/\delta g_{\mu\nu}$ in the integrand delivers the expectation value of the action variation, as suggested by our notation.  Expanding the previous variation of  the action to zeroth order in the fluctuations we just obtain the Einstein equations for the background,
\begin{equation}\label{eq:background Einstein}
	M_P^2 \, \bar{G}^{\mu\nu}=\bar{T}^{\mu\nu}.
\end{equation}
Because a vertex  linear in the fluctuations  cannot be part of a 1PI diagram,  quantum corrections to the previous equation result from terms that are at least quadratic in the fluctuations.  One of these corrections arises  from the expectation of the energy-momentum tensor of the fields in the given spacetime background,
\begin{equation}\label{eq:EMT exp}
	M_P^2 \, \bar{G}^{\mu\nu}=\bar{T}^{\mu\nu}+\langle \partial^\mu \chi \partial^\nu \chi -\frac{1}{2}\bar{g}^{\mu\nu}
	\left(\partial_\sigma \chi \partial^\sigma \chi+m_0^2 \chi^2\right)
	\rangle_{1PI},
\end{equation}
where indices are raised with the background metric. At this stage, the quantum corrected equations of motion resemble those of semiclassical gravity, in which the energy momentum tensor is replaced by its expectation (in our treatment, though, the  expectation would include radiative corrections with metric fluctuations running inside loops.)  In our specific setting,  this correction would account for the backreaction on cosmic expansion caused by the decay of the inflaton field, which we have ignored in this article. 

Expanding the Einstein tensor to second order in the metric fluctuations we would obtain the effective energy tensor of the metric fluctuations. We shall not write down the resulting expansion here, though it is clear that the latter will consist of quadratic terms in the fluctuations containing up to two derivatives of the metric. At short wavelengths, we expect the dominant terms to be captured by Isaacson's energy-momentum-tensor \cite{Isaacson:1968zza}, whose expectation value has been the central focus of this work. 

This framework offers a natural way to fix the counterterms needed to renormalize the divergences that appear in the different expectations, and thus fix the finite pieces of the counterterms that remained undetermined otherwise. Say, following the same  methods of Section \ref{sec:Regularization and Renormalization}, we can expand the spatial components of equation (\ref{eq:EMT exp}) in the number of time derivatives acting on the background fields. At zeroth order in derivatives, leaving out terms that vanish because of conditions (\ref{eq:mi cond}), we find
\begin{equation}\label{eq:EMT exp 0}
	\langle T^{ij}\rangle^{(0)}_{1PI}=\frac{\bar{g}^{ij}}{8(2\pi)^2}\sum_i \sigma_i m_i^4 \log \frac{2\Lambda}{a m_i},
\end{equation}
where the effective masses of the regulator fields are given by $m_i^2=M_i^2+\lambda\bar{\phi}^2.$
Clearly, the terms that diverge as the regulators are decoupled again renormalize the cosmological constant and the inflaton potential.  If we include the observed value of the cosmological constant in the background energy-momentum tensor of equation (\ref{eq:background Einstein}), it is thus natural to choose a counterterm that completely eliminates the contribution from (\ref{eq:EMT exp 0}). Such a counterterm is precisely that of equations (\ref{eq:counterterms}) and (\ref{eq:c0}), with the arbitrary finite piece in the latter set to zero. Similarly, equation (\ref{eq:EMT exp 0})  modifies  the effective energy density of the inflaton field by renormalizing its mass and quartic self-couplings. Demanding that the inflaton potential that appears in equation (\ref{eq:background Einstein})   is the  renormalized one, we are again led to the counterterm (\ref{eq:c0}) with the finite constant set to zero. 

There are no contributions with a single time derivative  in equation (\ref{eq:EMT exp}), and those with two derivatives are
\begin{equation}
\langle T^{ij}\rangle^{(2)}=-\frac{1}{a^4}\frac{\delta^{ij}}{12(2\pi)^2}
\left(\mathcal{H}^2+2\mathcal{H}'\right)
\sum_i  \sigma_i m_i^2 \log \frac{2\Lambda}{a\, m_i}
\end{equation}
This correction is proportional to $\bar{G}^{ij}=-a^{-4}(\mathcal{H}^2+2\mathcal{H}')\delta^{ij}$ and thus renormalizes the value of the Planck mass by an inflaton-dependent factor. If the constant $M_P$ that appears in equation (\ref{eq:background Einstein}) corresponds to the actually measured quantity, we thus need to subtract the whole quantum correction with an Einstein-Hilbert  counterterm (\ref{eq:counterterms}), with $c_2$ given by (\ref{eq:c2}) and its  finite piece set to zero. 

At this point it becomes clear that, in this context, a convenient renormalization scheme   involves the subtraction of the log divergent terms that appear in the different quantum corrections.  Such a prescription is somewhat analogous to minimal subtraction scheme often employed in field theories in flat spacetime. 

\end{fmffile}

\acknowledgments

It is a pleasure to thank Eugene A.~Lim and an anonymous referee for useful comments  on  earlier versions of this manuscript.

\end{document}